\documentclass[preprint2]{aastex}
\usepackage{mathptmx}
\usepackage{graphicx}
\usepackage{epstopdf}

\begin{document}

\title{Simulated Space Weathering of Fe- and Mg-rich Aqueously Altered Minerals Using Pulsed Laser Irradiation}
\author{Kaluna, H. M., Ishii, H. A., Bradley, J. P., Gillis-Davis, J. J., Lucey, P. G.}
\affil{Hawaii Institute of Geophysics and Planetology, University of Hawaii, Honolulu-HI-96822}
\email{kaluna@higp.hawaii.edu, hope.ishii@higp.hawaii.edu, jbradley@higp.hawaii.edu,
 gillis-davis@higp.hawaii.edu, lucey@higp.hawaii.edu }

\begin{abstract}

Simulated space weathering experiments on volatile-rich carbonaceous chondrites (CCs) have resulted in contrasting spectral behaviors (e.g. reddening vs bluing).  The aim of this work is to investigate the origin of these contrasting trends by simulating space weathering on a subset of minerals found in these meteorites.  We use pulsed laser irradiation to simulate micrometeorite impacts on aqueously altered minerals and observe their spectral and physical evolution as a function of irradiation time.  Irradiation of the mineral lizardite, a Mg-phyllosilicate, produces a small degree of reddening and darkening, but a pronounced reduction in band depths with increasing irradiation. In comparison, irradiation of an Fe-rich aqueously altered mineral assemblage composed of cronstedtite, pyrite and siderite, produces significant darkening and band depth suppression.  The spectral slopes of the Fe-rich assemblage initially redden then become bluer with increasing irradiation time.  Post-irradiation analyses of the Fe-rich assemblage using scanning and transmission electron microscopy reveal the presence of micron sized carbon-rich particles that contain notable fractions of nitrogen and oxygen.  Radiative transfer modeling of the Fe-rich assemblage suggests that nanometer sized metallic iron ({\it n}pFe$^0$) particles result in the initial spectral reddening of the samples, but the increasing production of micron sized carbon particles ($\mu$pC) results in the subsequent spectral bluing.  The presence of {\it n}pFe$^0$ and the possible catalytic nature of cronstedtite, an Fe-rich phyllosilicate, likely promotes the synthesis of these carbon-rich, organic-like compounds.  These experiments indicate that space weathering processes may enable organic synthesis reactions on the surfaces of volatile-rich asteroids.  Furthermore, Mg-rich and Fe-rich aqueously altered minerals are dominant at different phases of the aqueous alteration process.  Thus, the contrasting spectral slope evolution between the Fe- and Mg-rich samples in these experiments may indicate that space weathering trends of volatile-rich asteroids have a compositional dependency that could be used to determine the aqueous histories of asteroid parent bodies.

\end{abstract}


\section {Introduction}
Our ability to characterize the composition of Solar System bodies
greatly depends on remote sensing data and imaging
of spectral features.  However, space weathering processes, which alter the surfaces of 
airless bodies chemically, physically and optically, can introduce spectral variations
that reduce our ability to accurately assess the mineralogy of 
these bodies.  Both Solar System bodies that have been sampled by spacecraft, 
the Moon and the near Earth S-class asteroid Itokawa, 
show evidence of space weathering.  In particular, 
the sampled grains from both of these bodies have rims bearing {\it n}pFe$^0$ particles, 
which form in response to processes such as micrometeorite impacts and 
solar wind sputtering \cite[e.g][]{Conel:1970,Keller:1993,Noguchi:2011}.
Various types of large, micron-sized glassy agglutinates, which form during micrometeorite bombardment, 
are also observed in these samples \citep{McKay:1991,Ogliore:2015}.  
Both of these optically active space weathering products result in the reduction
of albedo, increase in spectral slopes (reddening), and the reduction of absorption features
with increasing exposure age \cite[see][and references therein]{Pieters:2000,Hapke:2001}.  

While the {\it n}pFe$^0$ and agglutinates observed in lunar soils and Itokawa
samples bear similarities, there are distinct differences in the particles
that are attributed to the differing gravities \citep{Ogliore:2015} and compositions of the 
two bodies.  For example, in contrast to lunar soils, a notable fraction of 
{\it n}pFe$^0$ particles are much less abundant on Itokawa and a notable fraction of these
particles are sulfur bearing \citep{Noguchi:2011}.  
The differences in these space weathering products may be
further amplified in the main asteroid belt, where the reduction in 
solar wind flux and micrometeorite impact energies 
\citep{Chapman:2002} may reduce the effectiveness 
of {\it n}pFe$^0$ and agglutinate production.  Nonetheless, 
spectral reddening has been observed among silicate-rich S-complex, 
and primitive, low-albedo C-complex main belt asteroids
\cite[e.g.][]{Jedicke:2004,Nesvorny:2005,Lazzarin:2006,Kaluna:2016,Fornasier:2016}.
Although spacecraft have not yet sampled main belt asteroids and we are limited in our 
understanding of space weathering products on these bodies,
meteorites and meteorite analogs provide useful tools that enable the 
study of space weathering processes through experimentation.  
  
Most published space weathering studies have focused on the spectral 
evolution of anhydrous minerals
found in lunar soils and ordinary chondrites (OC), the meteorite analogs 
of S-complex asteroids \cite[e.g.][]{Hapke:1966,Yamada:1999,Sasaki:2001}.
In contrast, the number of experiments on
CC meteorites is relatively small (Table \ref{tab_pub_sw_results}) 
compared to the large compositional diversity of CC meteorites \citep{Burbine:2002}.
Ion or pulsed laser irradiation of some CCs results in 
the reddening and darkening of meteorite spectra
similar to the optical variations observed among experimentally space 
weathered OCs.  However, experiments on the aqueously altered CC meteorites
Tagish Lake, Mighei and Murchison reveal 
a new trend of decreased spectral slopes (i.e. flattening or bluing) with increased 
exposure to ion bombardment or laser irradiation
\citep{Hiroi:2004,Vernazza:2013,Matsuoka:2015,Lantz:2015}.

In comparison, both reddening and bluing trends have been observed 
among C-complex asteroids.  However, \cite{Lazzarin:2006} and \cite{Kaluna:2016} 
suggest the bluing trend found by \cite{Nesvorny:2005} 
arises from compositional variations rather than space weathering processes.  
Observations of the C-complex, Themis asteroid family and its younger 
sub-family, the Beagle asteroids, show spectral reddening with 
increasing age \citep{Kaluna:2016,Fornasier:2016}.  
Due to their common origin \citep{Hirayama:1918,Nesvorny:2008} 
and similarity in spectral features \citep{Kaluna:2015,Kaluna:2016a},
the observed spectral reddening among these two families is unlikely the result of 
compositional heterogeneity \citep{Kaluna:2016}.
Though the Themis and Beagle asteroid families provide a useful tool to study space weathering, 
they are currently the only known C-complex asteroids that originate from the same parent body 
while having different dynamical (i.e. surface) ages.  
Thus, the bluing observed in simulated space 
weathering experiments may still be representative of space weathering 
trends on other C-complex asteroid families.

The volatile-rich CI and CM meteorites contain a significant fraction of
phyllosilicates (up to 80 wt.\% \cite{Cloutis:2011}), which are the group of minerals 
associated with the spectral features in C-complex asteroids \cite[e.g.][]{Lebofsky:1981,Vilas:1989}.  
Modal mineralogy of volatile-rich CCs suggests that Fe-rich phyllosilicates give way to 
Mg-rich end members as aqueous alteration 
progresses \citep{Howard:2009,Howard:2011,Rubin:2007a}. 
In this study, we have selected a subset of the minerals and assemblages 
that are representative of these two end phases of aqueous
alteration to investigate the role these minerals have
on the space weathering trends of volatile-rich CCs and C-complex asteroids.   
We performed pulsed-laser irradiation on the Mg-rich phyllosilicate lizardite and 
an Fe-rich assemblage consisting of cronstedtite, pyrite and siderite 
to simulate the effects of micrometeorite impacts.  
The same experimental procedure was used with San Carlos olivine to 
compare the spectral trends of our aqueously altered samples with those of lunar and 
S-type mineralogies.

\section{Methodology}
\subsection {Sample characterization and preparation}
For this study we irradiated two sets of aqueously altered terrestrial minerals, 
the Mg-rich phyllosilicate lizardite
(Mg$_3$(Si$_2$O$_5$)(OH)$_4$), and an Fe-rich assemblage composed of 
cronstedtite (Fe$^{2+}$$_2$Fe$^{3+}$(SiFe$^{3+}$)O$_5$(OH)$_4$), 
pyrite (FeS$_2$) and siderite (FeCO$_3$).  
The lizardite was manually extracted from an assemblage where 
lizardite sheets were interspersed with artinite (Mg$_2$(CO$_3$)(OH)$_2$á3H$_2$O)
needles.  The Fe-rich assemblage, which occurred as siderite 
and pyrite crystals embedded in a cronstedtite matrix, 
was isolated from a basalt-like host rock using a rotary tool and diamond tipped bits.

\begin{figure*}[h!]
\begin{center}
\includegraphics[width=0.5\linewidth,angle=0]{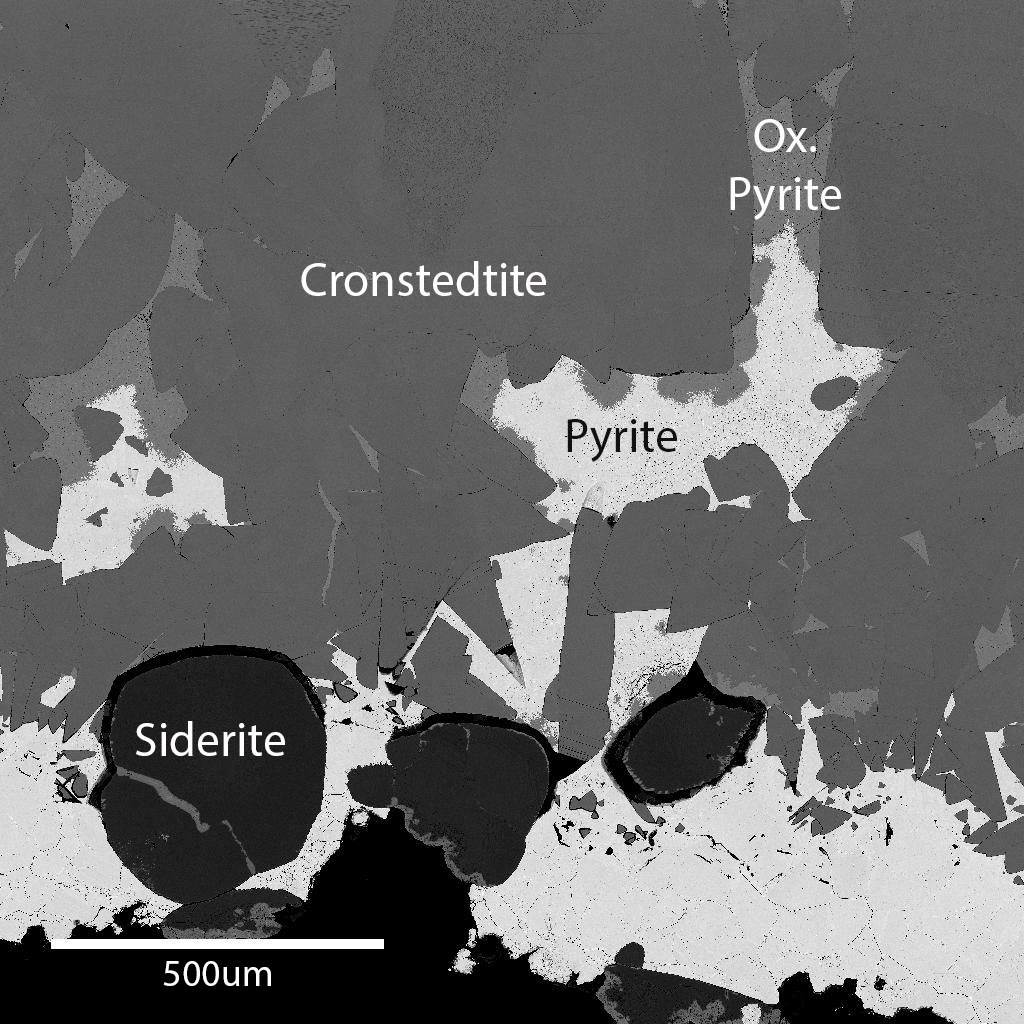} 
\end{center}
\caption[Electron microprobe image of fe-rich sample]{\label{eprobe-sei}
Secondary electron image of the mineral phases in the 
Fe-rich assemblage taken with the HIGP electron microprobe.   
In addition to cronstedtite, pyrite, and siderite, oxidized pyrite (ox. pyrite) was 
observed near pyrite and cronstedtite interfaces. }
\end{figure*}

Chemical compositions of lizardite and Fe-rich assemblage thin sections 
were determined using the Hawaii Institute of Geophysics and Planetology's (HIGP)
electron microprobe.
The electron microprobe is a JEOL JXA-8500F field-emission gun microanalyzer 
with 5 wavelength-dispersive spectrometers. 
Beam currents of 4.0 nA for lizardite and 5.0 nA for the Fe-rich assemblage were used 
with a spot size of 10$\mu$m for both samples.
For the Fe-rich assemblage, Fe-olivine was used as a mineral standard for Si, 
San Carlos olivine for Mg, anorthite for Al, and troilite for Fe and S with errors of $\sim$1\%.  
The mineral standards used for lizardite were San Carlos olivine for Si, Fe and Mg,
and plagioclase for Al, with errors of ranging from $\sim$0.3-7\%. 
A secondary electron (SE) image of the cronstedtite, pyrite and siderite phases
in the Fe-rich assemblage is shown in Figure \ref{eprobe-sei}.  
Reactions near the pyrite and cronstedtite interfaces have oxidized 
some of the pyrite, resulting in leached sulfur and an
enrichment in iron and oxygen.
The fractional abundances for lizardite and each of the phases in the Fe-rich
assemblage are given in Table \ref{tab_microprobe}.  

To determine the modal mineralogy of the samples, we used x-ray diffraction.  
Powdered samples were measured using the HIGP's InXitu Inc. 
X-ray diffraction~/~X-ray fluoresence (XRD/XRF) instrument. 
The XRD analysis revealed that the lizardite sample was pure and the Fe-rich assemblage
was composed of $\sim$73 wt.\% cronstedtite, 23 wt.\% pyrite and 
3 wt.\% siderite.  

\subsection{Experimental procedure}
The powdered samples were dry-sieved into size fractions greater and 
less than 75 $\mu$m.  The laser irradiation experiments utilized $\sim$0.5 
gram of each sample from the $<$75 $\mu$m size fraction powders, which were 
placed into a glass beaker as a loose powder. 
To simulate space weathering by 1~$\mu$m dust impacts with velocities typical 
at 1 AU, we used a pulsed laser with a focused spot size of $\sim$0.25~mm, a pulse
energy of 30 mJ \citep{Sasaki:2001,Sasaki:2003} 
and pulse duration of 6-8 nanoseconds.  The pulse duration 
is similar to the timescale of micrometeorite dust impacts \citep{Yamada:1999}.
The samples were irradiated under vacuum at pressures of 10$^{-5}$ to 10$^{-6}$ mbar
with a Continuum Surelite-I-20 Nd:YAG 1064 nm laser pulsed at a frequency of 20 Hz.
A manually adjusted pan/tilt mirror was used to move the laser spot and irradiate the 
full area of the sample.  The incident laser gardened the samples vertically during 
irradiation, leading to uniform simulated weathering. 

Three sets of experiments were conducted on each sample to quantify variability 
in the space weathering trends.  Each of the samples were irradiated multiple 
times over a consistent set of intervals in order to
monitor the spectral evolution of the samples
as a function of irradiation time.  The lizardite and olivine samples 
were irradiated at intervals of 
2.5, 2.5, 5.0, 5.0, 5.0, 10.0, 10.0, 10.0 and 10.0 minutes for a total of 60 minutes.  
The irradiation intervals for Fe-rich assemblage followed the same interval sequence 
as olivine and lizardite, however the sequence was halted at 40 minutes.  We 
were forced to shorten the irradiation sequence due to the 
the considerable volatility of the Fe-rich samples in response to the 
laser, which resulted in a notable portion of the samples being displaced 
from the glass beakers into the vacuum chamber.  
Gas partial pressures of water and other volatiles released during irradiation
were monitored using a 100amu Residual 
Gas Analyzer (RGA) from Stanford Research System.    
The RGA is attached to the vacuum chamber, and uses an ionizer
and a quadrapole mass filter to measure volatiles released 
as a function of atomic mass.
  
Spectral data were obtained using a visible and near-infrared wavelength (0.35-2.5~$\mu$m) 
FieldSpec 4 Spectroradiometer from Analytical Spectral Devices Inc.  
The spectroradiometer has resolutions of 3~nm and 10~nm in the 0.35-1.0~$\mu$m and 
the 1.0-2.5~$\mu$m regions, respectively.  
The reflectance spectra were measured relative to a 99\% reflectance 
LabSphere Spectralon Standard
and taken at incidence and emission angles
of {\it i}=30$^\circ$ and {\it e}=0$^\circ$.  The spectral measurements were 
conducted outside of the vacuum 
chamber and made after each irradiation interval.  
Samples were prepared for spectral measurements by placing them in a 
sample holder and leveling the sample surface by moving a flat spatula across the top of the 
holder, taking care not to compress the sample.  
Spectra were taken at four positions on the sample, after which the sample was removed, 
mixed and placed back into the holder, and the sequence repeated twice more. This process 
results in 12 spectra that account for packing effects and sample variability during 
placement into the holder.  

\subsection{Post-irradiation sample characterization} 
To characterize the irradiation products of the lizardite and Fe-rich assemblage
samples, we obtained nanometer scale, high resolution images
using the Pacific Biosciences Research Center's (PBRC) 
Hitachi S-4800 field emission scanning electron microscope (SEM).  
SE images were taken of a 60-minute irradiated lizardite sample and 
a 40-minute irradiated Fe-rich assemblage.  
Prior to imaging, the irradiated mineral grains were 
mounted onto sample stubs with double sided carbon tape.
The SE images of lizardite were obtained using a current of 5kV and acceleration 
voltages of 6.2-10.9$\mu$A, whereas a configuration of 2kV and 8.6-16.8$\mu$m was used 
for the Fe-rich assemblage. 

Two melt grains from the irradiated Fe-rich samples were selected for further 
characterization and made into Focused Ion Beam (FIB) sections for
analysis with transmission electron microscopy (TEM). 
The first grain was cut and prepared using an FEI Strata 235 dual beam FIB
at the Molecular Foundry, Lawrence Berkeley Lab.  The second FIB section 
was prepared with the FEI Helios 650 dual beam FIB 
at Oregon State University's Electron Microscopy Facility.   
A focused Ga+ ion beam was used for site-selective extraction of 
laser-irradiated grain sections 
and were attached to Cu half-grids for subsequent TEM analyses.  
Field emission SE and secondary ion images were obtained of the 
FIB sections during sample preparation, 
and standard FIB sample preparation procedures were applied with final 
thinning of the sections on the TEM grid \cite[e.g.][]{Ishii:2010}.

The Fe-rich assemblage FIB sections were analyzed by 
scanning transmission electron microscopy on 
two FEI Titan (S)TEMs. One microscope is a 80-200 kV Titan at 
Oregon State University (OSU), 
and the other is a 60-300kV Titan (referred to as TitanX) at the National Center for 
Electron Microscopy (NCEM), Molecular Foundry, Lawrence Berkeley Lab.  
Each instrument has a high brightness FEG and ChemiSTEM technology consisting 
of four Bruker silicon drift detectors (SDDs) for ~0.7 sr total solid angle acceptance and 
140 eV energy resolution at Mn K-alpha.  
Energy dispersive x-ray (EDX) mapping on all FIB sections was 
carried out with 10 eV/pixel dispersion. The convergence semi-angle alpha 
was 10.0 mrad on both instruments, and the HAADF inner semi-angle beta was set at 
40 mrad at OSU and 63 mrad at NCEM.
EDX mapping on the first FIB section 
was carried out on the OSU Titan at 200 kV and 400 pA with 
pixels of 6 nm on a side, 998 pixels width and 662 pixels height, with a 
total collection time of 4.3 ms/pixel.  
Full x-ray fluorescence spectra were collected at each pixel, however 
due to its superior low-energy performance, additional chemical analyses
were carried out on the NCEM TitanX.  EDX mapping on 
the second FIB section was also conducted with the NCEM TitanX using 200kV and 
full x-ray fluorescence spectra collected at each pixel.
The map was carried out at 900 pA beam current with pixels
of 9.3 nm on a side, 352 pixels wide and 292 pixels high, with a total collection 
time of 8.2 ms/pixel.  
Maps were processed using Bruker's Esprit 1.9 analysis software, and 
regions of interest within the FIB sections were drawn and spectra from pixels within 
those regions summed, fitted and quantified to obtain chemical compositions.  

\subsection{Radiative Transfer Modeling}
We used a modification of the \cite{Hapke:2001} radiative transfer model from 
\cite{Lucey:2011} to estimate the abundance and 
model the optical effects of nanophase and microphase metallic 
iron particles ({\it n}pFe$^0$ and $\mu$pFe$^0$,
respectively) on the irradiated sample spectra of olivine, lizardite and the Fe-rich
assemblage samples.
The model uses Mie theory to look at the effects of reddening 
and darkening from nanophase 
particles ($<$ 50~nm), in addition to the purely darkening 
effect of microphase particles ($>$ 50~nm).  
SEM imaging and TEM analysis revealed the presence of micron sized 
carbon particles on the irradiated Fe-rich sample grains, so we 
also modeled the effects of nanophase and microphase amorphous carbon 
({\it n}pC and $\mu$pC respectively) particles in addition to 
{\it n}pFe$^0$ and $\mu$pFe$^0$ for the Fe-rich samples.

The spectra and associated viewing geometry ($i$=30$^\circ$ and $e$=0$^\circ$) of our fresh 
(non-irradiated) samples are used as input into the model, along with
the mean grain diameters (37.5$\mu$m), grain densities ($\rho_{olivine}$ = 3.32g/cm$^3$,
$\rho_{lizardite}$ = 2.55g/cm$^3$ and $\rho_{Fe-rich}$ = 3.34g/cm$^3$) and 
real indices of refraction ({\it n}$_{olivine}$ = 1.64,
{\it n}$_{lizardite}$ = 1.60 and {\it n}$_{Fe-rich}$ = 1.72).  
Because it is the most abundant mineral in the sample, 
we use the real index of refraction and density of cronstedtite as 
input into the model for the Fe-rich assemblage.  The optical constants
used in our models are derived from \cite{Cahill:2012} for iron and 
\cite{Rouleau:1991} for the amorphous carbon.

These initial parameters are first used to 
obtain the complex index of refraction spectrum for each sample, which is then used 
along with estimated values of the nanophase and microphase particle abundances
to derive a modeled irradiated spectrum.  
For each spectrum, we use 
the IDL routine MPFIT \citep{Markwardt:2009}, to vary the input abundances and perform a 
non-linear least squares fit between our model and the observed spectrum.  
This allows us to derive robust model spectra and particle abundances
for each sample as a function of laser irradiation time.  

\section{Results}
\subsection{Spectroscopic Trends}
\begin{figure*}[h!]
\begin{center}$
\begin{array}{llllll}
\includegraphics[width=0.46\textwidth,angle=0]{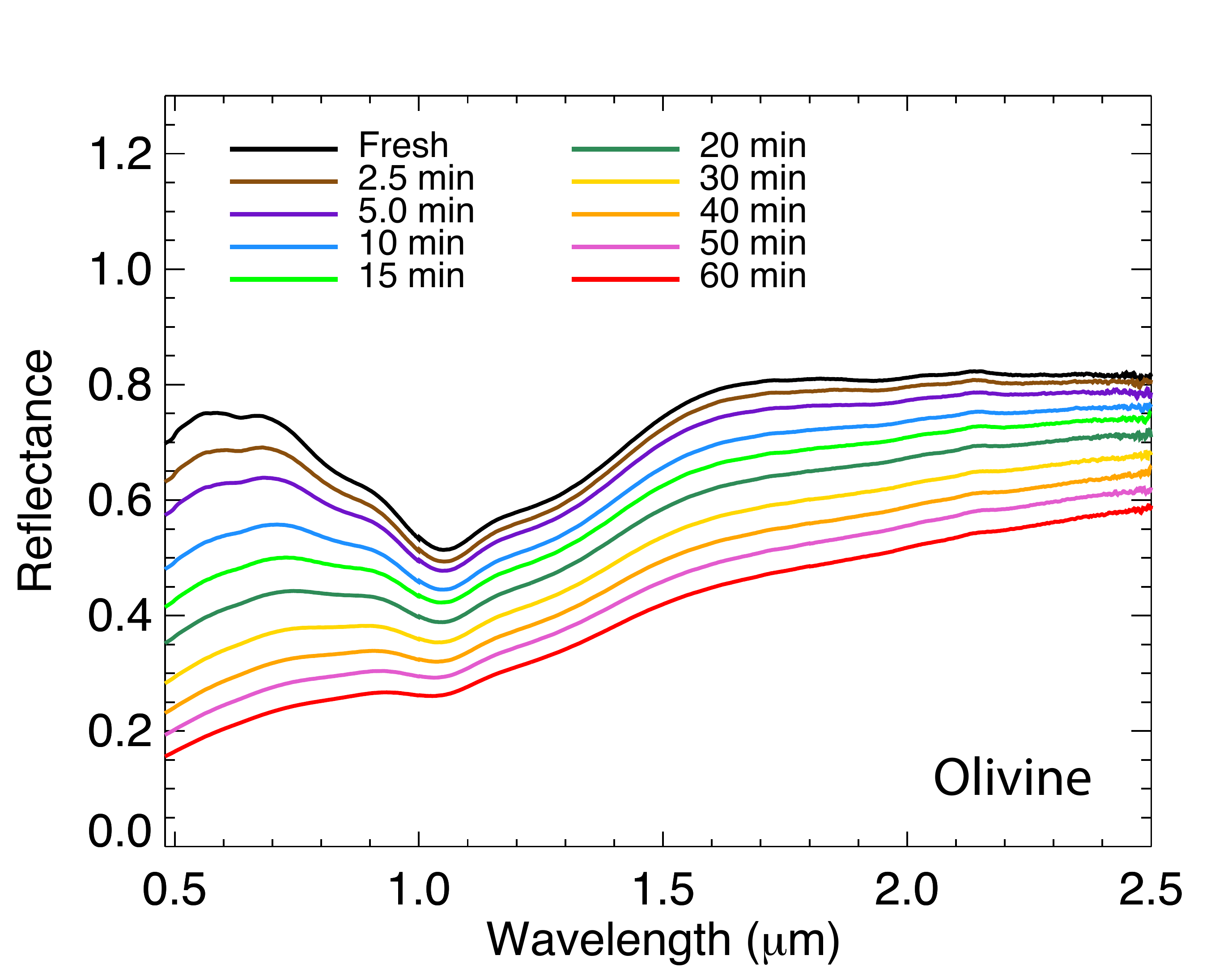} 
\includegraphics[width=0.46\textwidth,angle=0]{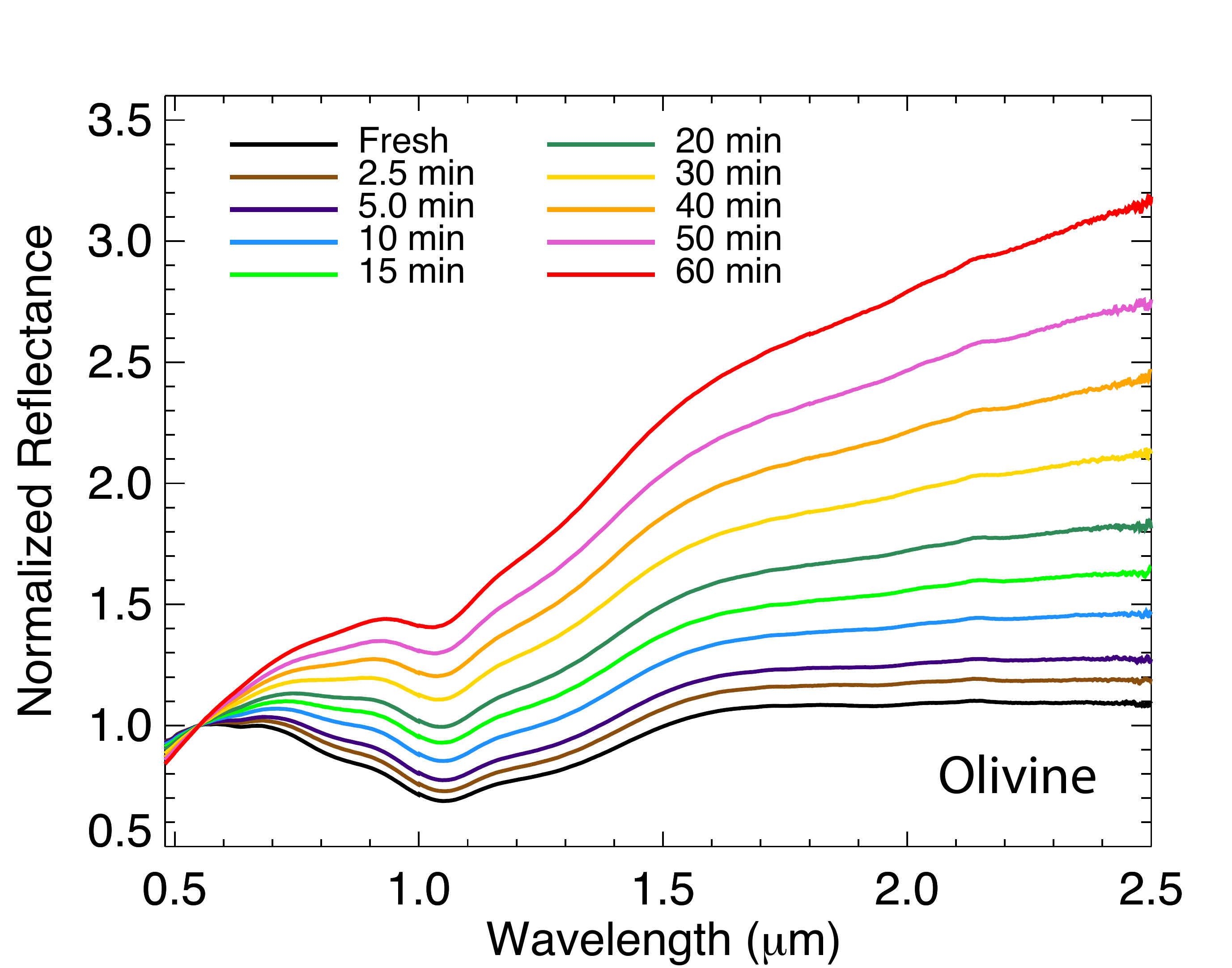}\\
\includegraphics[width=0.46\textwidth,angle=0]{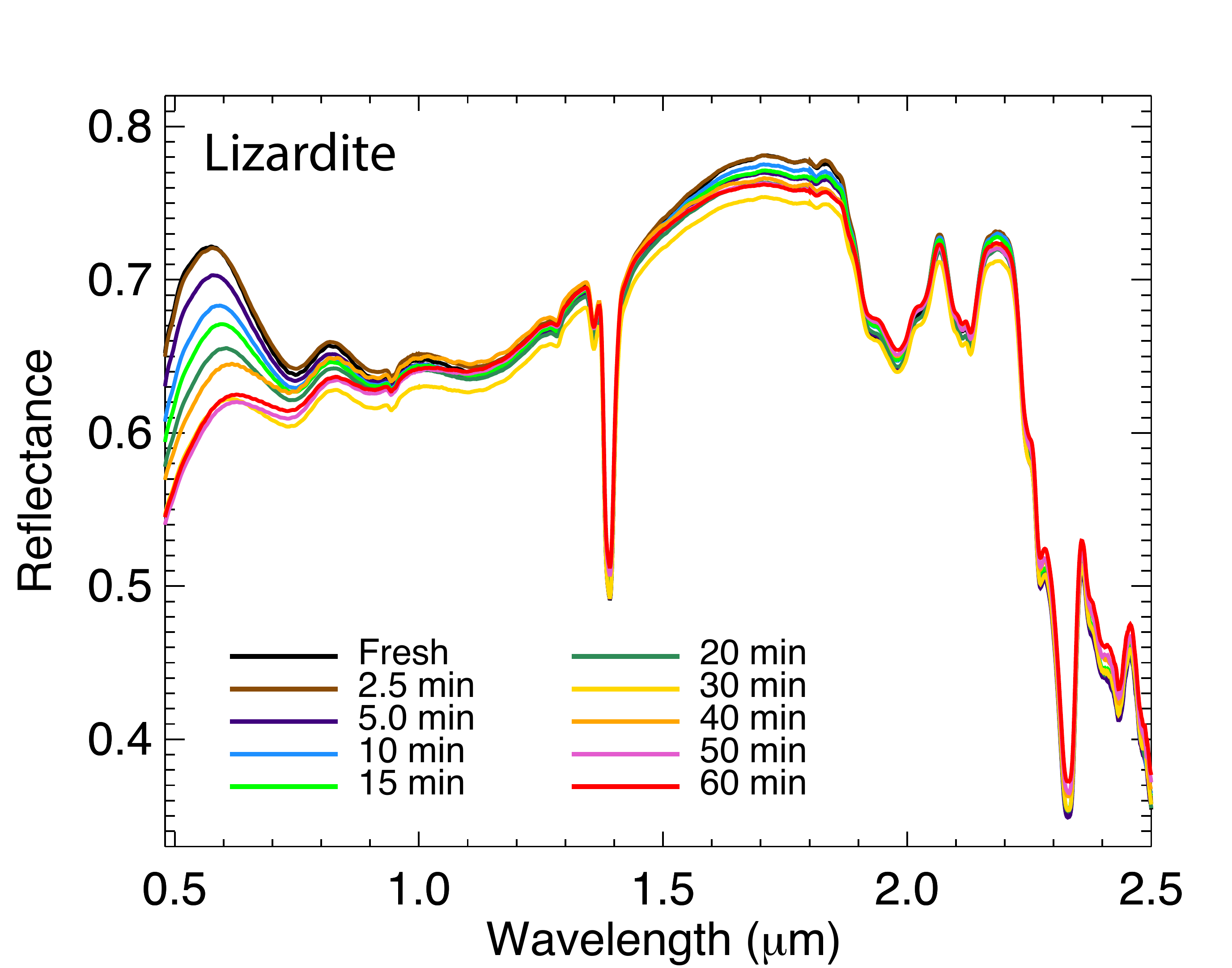} 
\includegraphics[width=0.46\textwidth,angle=0]{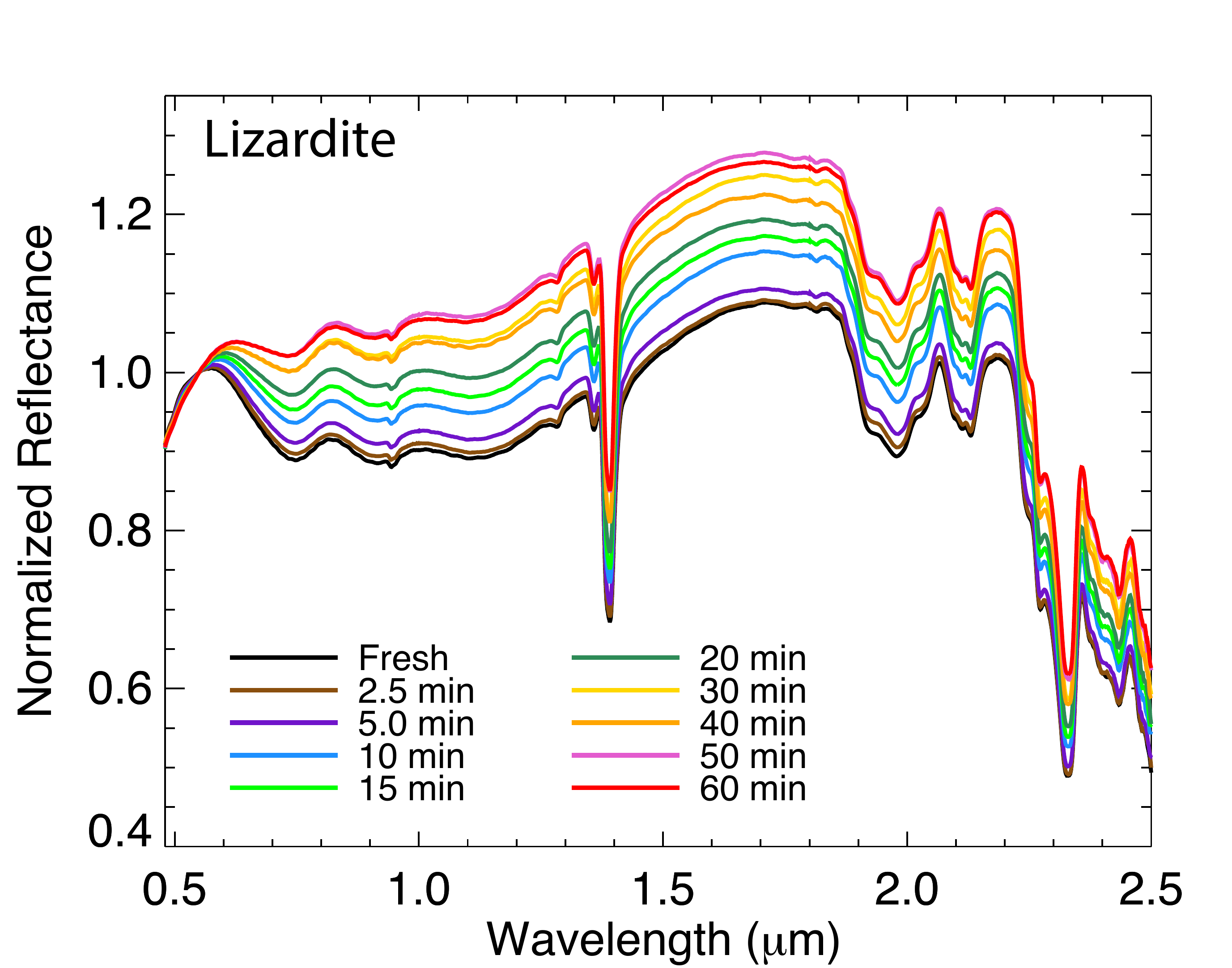} \\
\includegraphics[width=0.46\textwidth,angle=0]{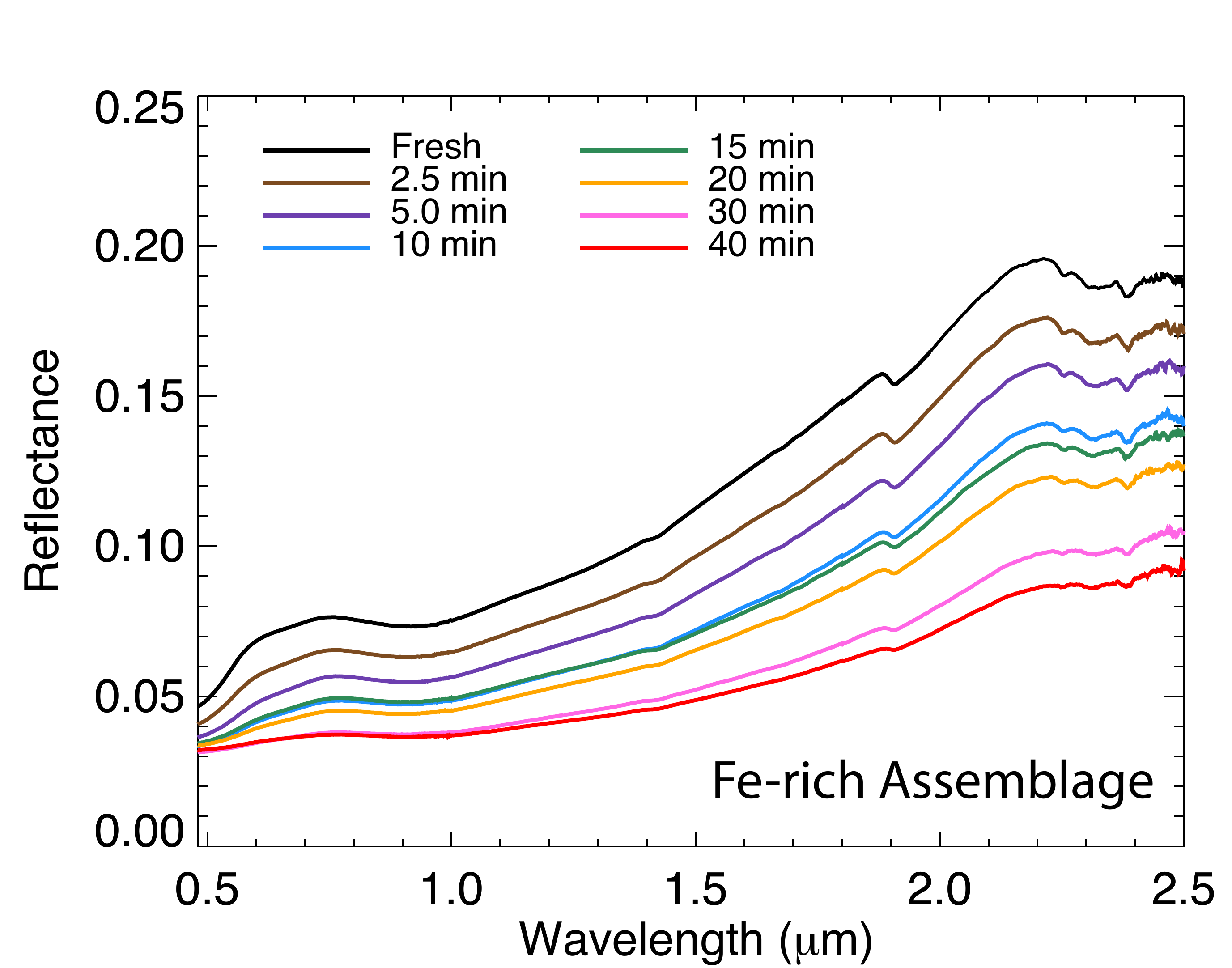} 
\includegraphics[width=0.46\textwidth,angle=0]{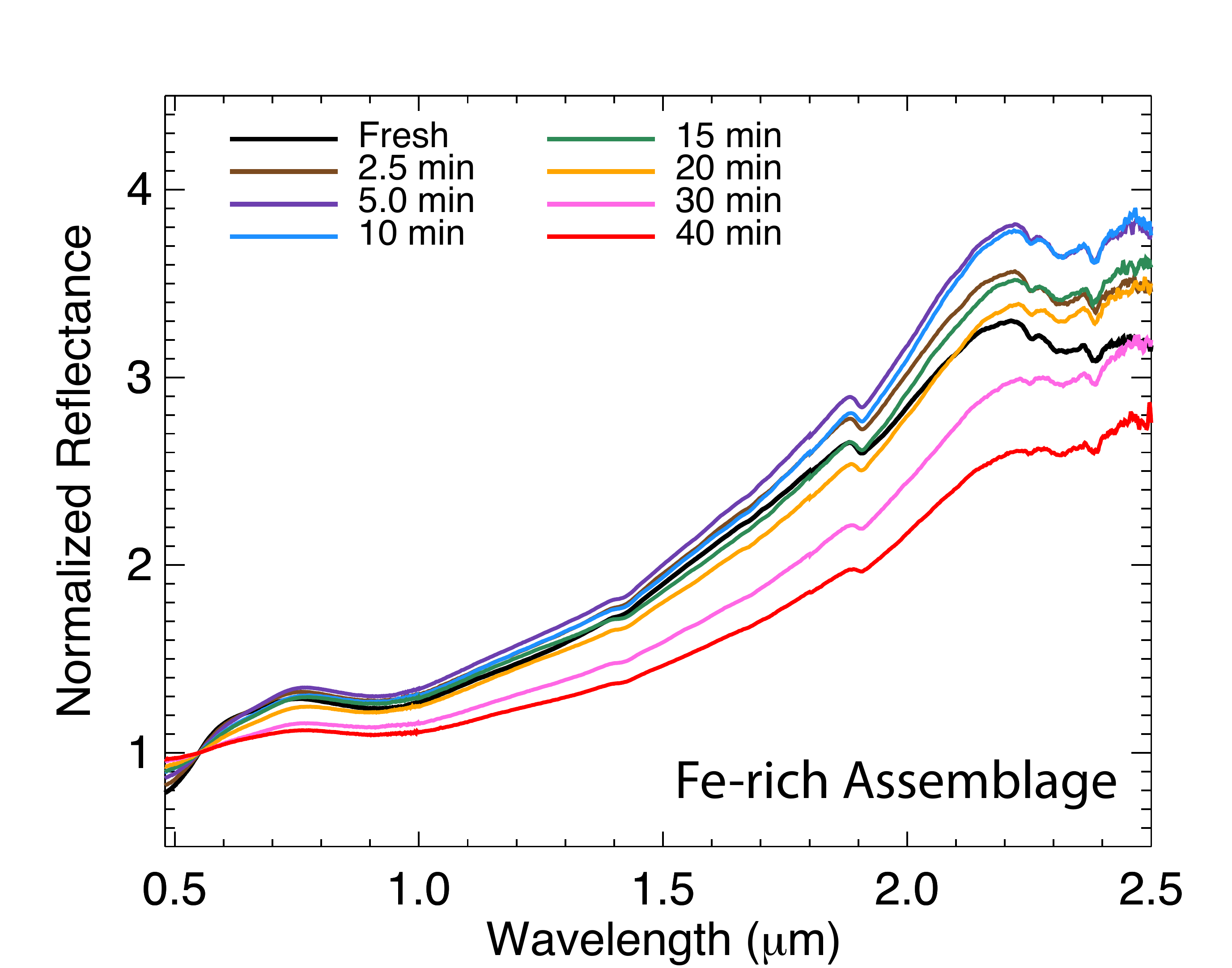}  \\
\end{array}$
\end{center}
\caption[Olivine, lizardite and Fe-rich assemblage reflectance and normalized spectra 
as a function of irradation time]{\label{sample_spectra}  
{\it Left:} Reflectance spectra for each sample as a function of laser irradiation time.  
The full 60 minute irradiation series are shown for olivine and lizardite, but due 
to sample loss in the Fe-rich samples, the total irradiation time was halted 
at 40 minutes.  {\it Right:} Reflectance spectra normalized to 0.55~$\mu$m.  }
\end{figure*}

\begin{figure*}[t!]
\begin{center}$
\begin{array}{ll}
\includegraphics[width=0.4\linewidth,angle=90]{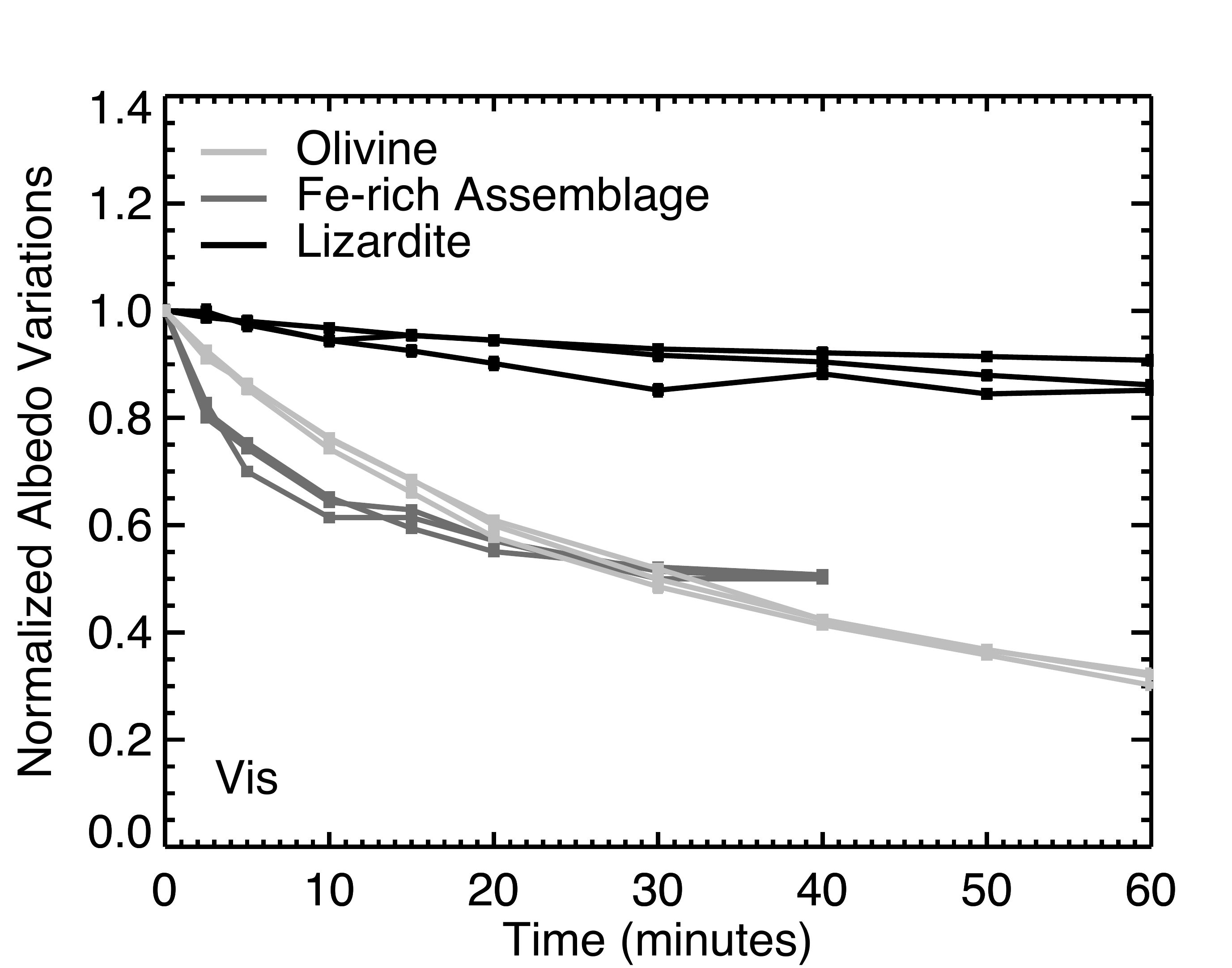} 
\includegraphics[width=0.4\linewidth,angle=90]{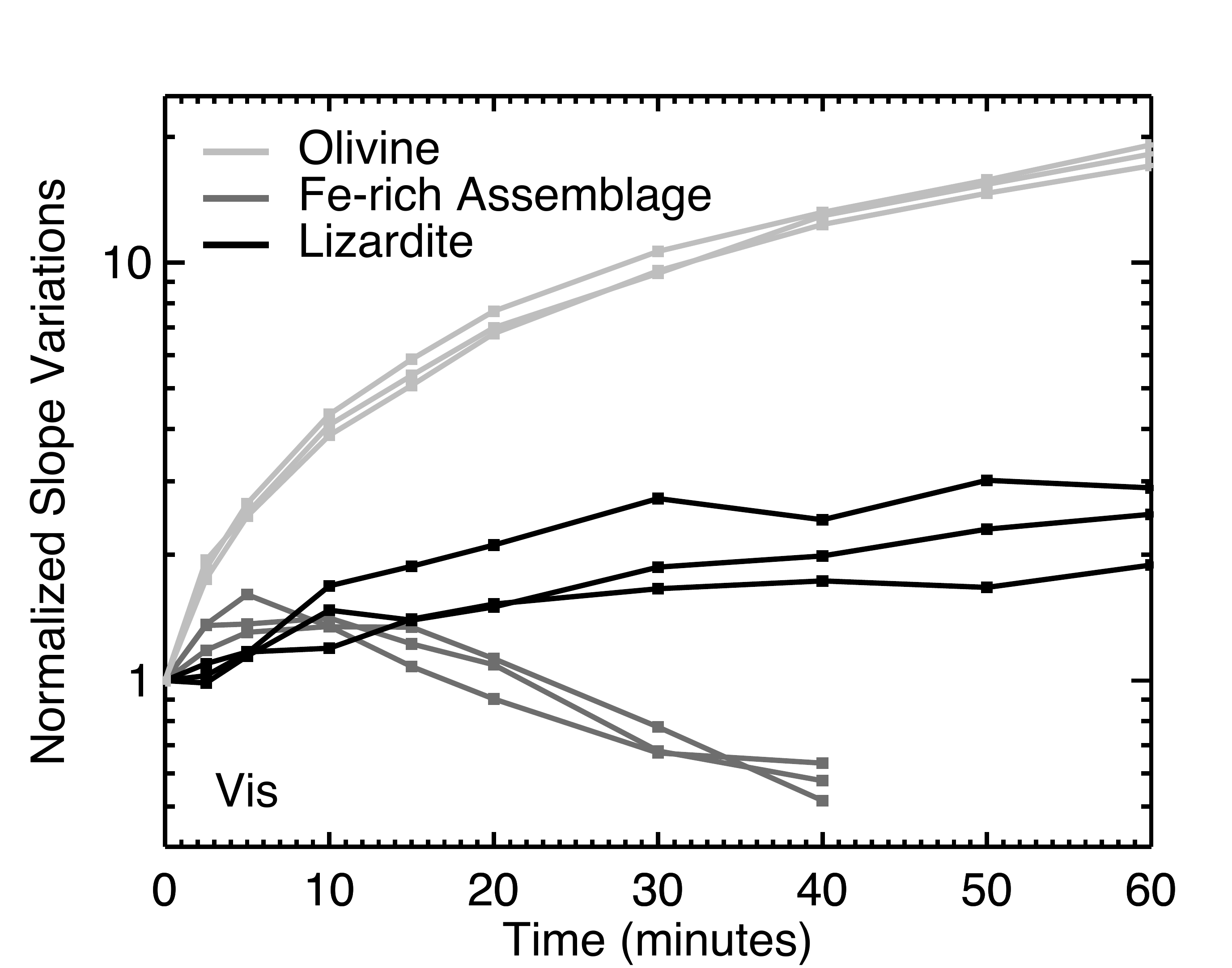} \\
\includegraphics[width=0.4\linewidth,angle=90]{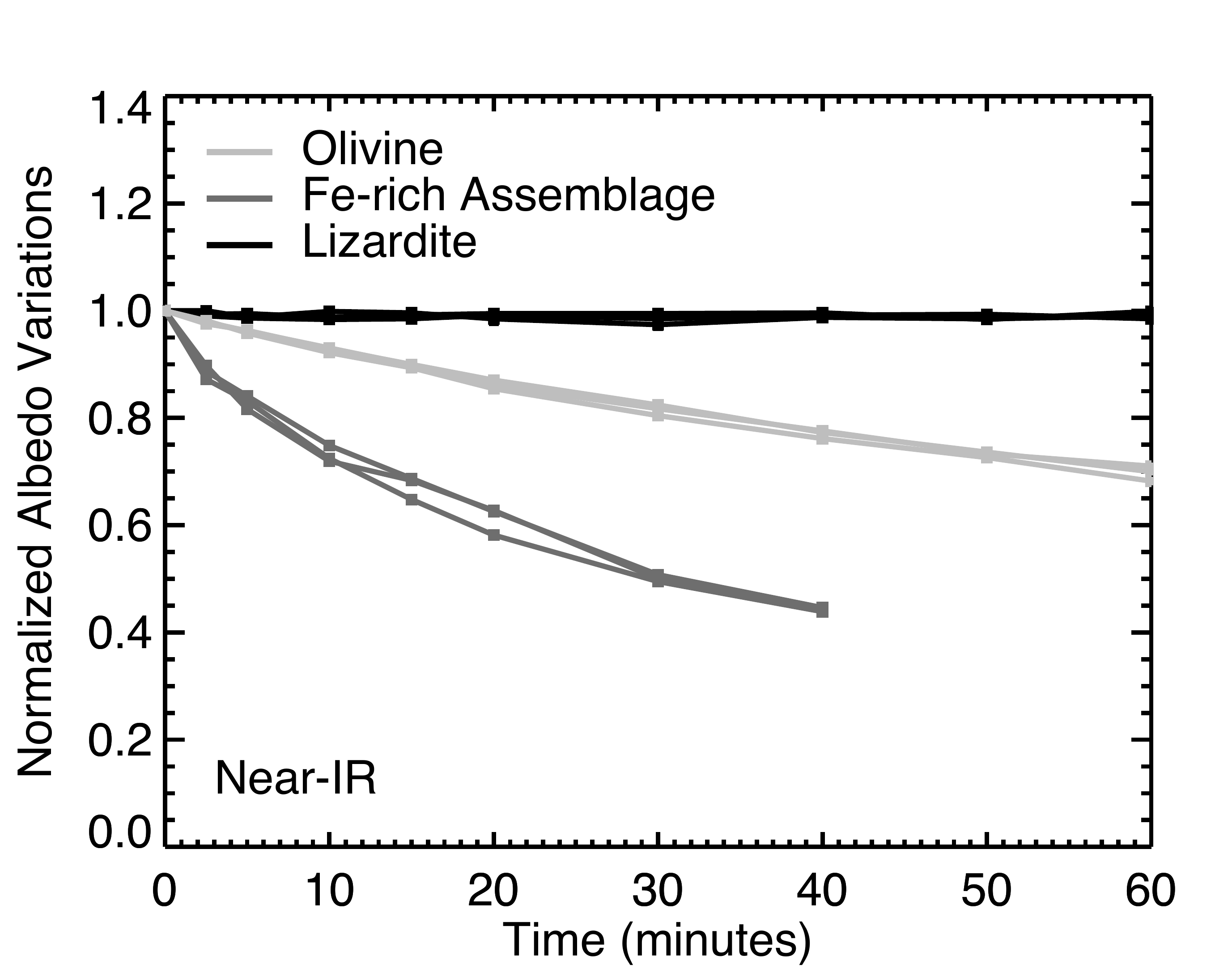} 
\includegraphics[width=0.4\linewidth,angle=90]{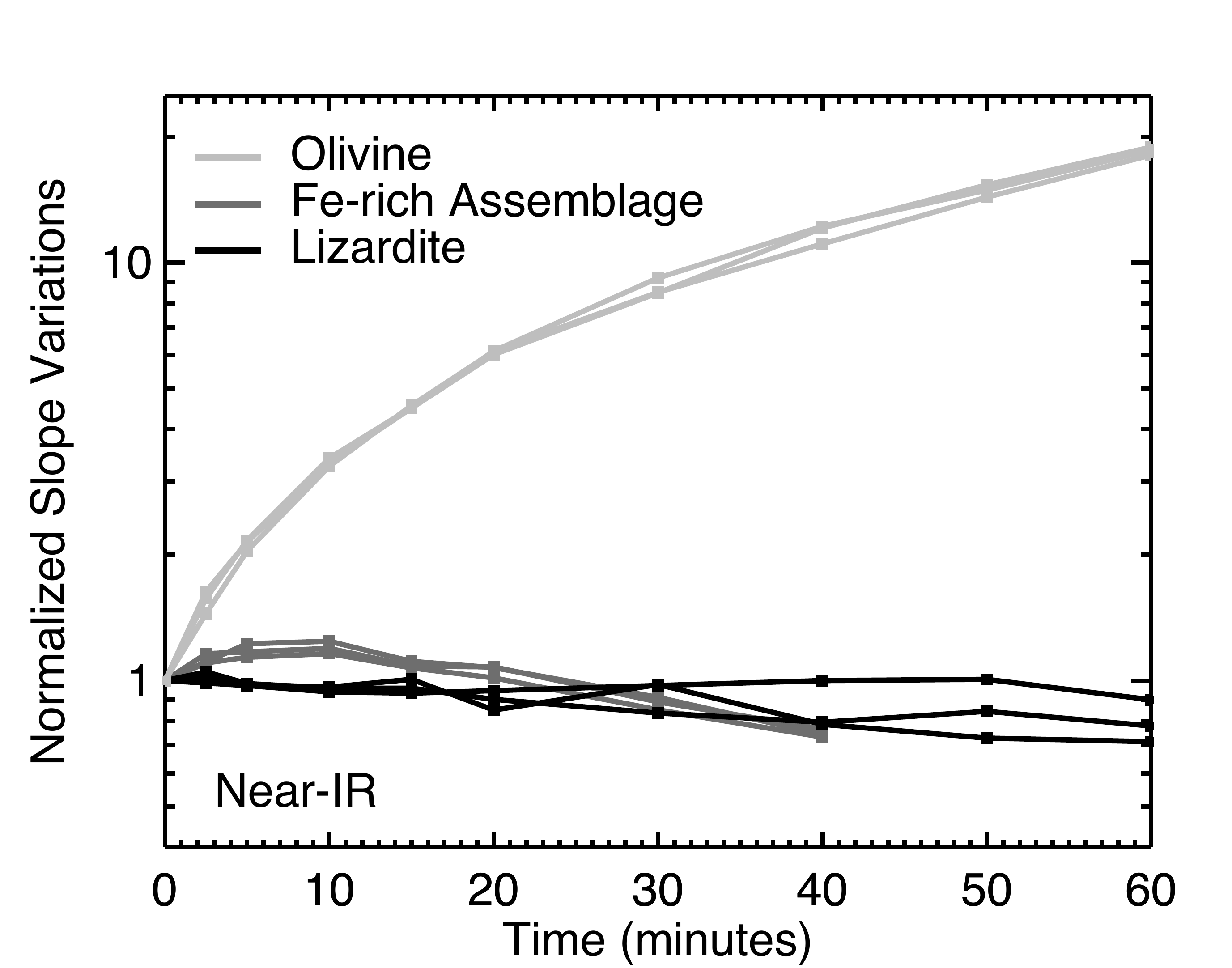} 
\end{array}$
\end{center}
\caption{\label{spec_compare}  
{\it Left:} Plots of relative slope changes in the visible (top) and 
NIR (bottom) for each experiment. 
Data have been normalized to the slope value of the fresh spectrum
for each experiment.   Visible slopes were measured using one continuum 
point in the visible when possible. 
{\it Right:} The relative changes in albedo for each experiment. 
Data have been normalized to the albedo value of the fresh spectrum
for each experiment.}
\end{figure*}

Spectra of olivine, lizardite and the Fe-rich assemblage as a function of irradiation time are
shown in Fig. \ref{sample_spectra}.  Although we use loosely packed powders in 
our experiments, our irradiated olivine shows a similar increase in spectral slope, 
and decrease in reflectance (i.e. albedo) and band depth when compared to 
other studies that use compressed pellets 
\cite[e.g][]{Yamada:1999,Brunetto:2005}. 
Figure \ref{sample_spectra} reveals that both the lizardite and Fe-rich 
samples exhibit irradiation dependent spectral variations. 
A comparison of the relative (data were normalized to the initial value 
in each experiment) albedo and slope variations 
of all samples is given in Fig. \ref{spec_compare}. 

Albedo and slope were defined by using visible and 
near-infared (near-IR) continuum points
not affected by absorption features.  Thus, due to the difference in mineral features,
the wavelengths of the continuum points used in our analysis are different for each sample.  
The reflectance values at 0.70~$\mu$m were used to measure the 
visible region albedo variations in olivine, and values at
0.57~$\mu$m and 0.60~$\mu$m were used for lizardite and 
the Fe-rich assemblage, respectively.  For the near-IR albedo variations, reflectance 
values at 2.30~$\mu$m for olivine, 2.16~$\mu$m for 
lizardite, and 2.20~$\mu$m for the Fe-rich assemblage were used. 

The wavelength value of 0.55~$\mu$m is often used to scale reflectance spectra in 
both experimental and observational data 
\cite[e.g.][]{Sasaki:2001,Bus:2002,Fornasier:2014}, 
so for consistency and to allow for 
comparison with other works, we used reflectance values at 0.55~$\mu$m to 
characterize slope variations in our sample spectra.
Due to broad absorption features and 
the lack of extended continuum regions in the visible, 
the visible slopes for these minerals were 
measured using one point in the visible and the other in the near-IR.  A linear fit to the 
normalized reflectance values at 0.70~$\mu$m and 1.60~$\mu$m were used to characterize
visible slope variations for olivine, and 0.57~$\mu$m and 1.70~$\mu$m were used for 
lizardite.  In contrast to the other minerals, the Fe-rich assemblage has an 
extended continuum region in the visible, so slope variations were measured using
a linear fit to continuum points at 0.60~$\mu$m and 0.75~$\mu$m.  
For the near-IR region, slopes 
were computed using normalized reflectance values at 1.60~$\mu$m and 2.30~$\mu$m
for olivine, and at 1.70~$\mu$m and 2.16~$\mu$m for lizardite.  The Fe-rich assemblage
has only one continuum point in the near-IR because of absorption features, 
so near-IR slopes were measured using one point in the visible (0.75~$\mu$m) 
and one point in the near-IR (2.20~$\mu$m).

Figure \ref{spec_compare} reveals that the evolution of spectral slopes is related to 
the wavelength dependency of the albedo variations. 
For instance, the reduction of albedo in olivine is greatest at shorter wavelengths, 
resulting in an increase of spectral slopes (reddening).  
Lizardite shows a slight decrease in 
albedo at visible wavelengths, but no variations in the near-IR, resulting in 
a slight reddening as a function of irradiation time.  
Although the variations are much smaller, the spectral behavior of lizardite is similar to olivine,
and is consistent with the production of {\it n}pFe$^0$ particles, 
which induce spectral reddening due to a decrease in the absorption 
efficiency of the {\it n}pFe$^0$ coated grains at longer wavelengths \cite{Hapke:2001}.  

In contrast to olivine and lizardite, the slopes of the 
Fe-rich assemblage samples initially redden, then reverse and  
become bluer with increased irradiation. 
As seen in Figure \ref{spec_compare}, the albedo of the Fe-rich assemblage 
initially decreases more rapidly at visible wavelengths than in the near-IR, but 
eventually begins to plateau while the near-IR albedo continues to decrease.  
This relative difference in the visible and near-IR albedo trends results in the 
reversal from slope reddening to bluing.  Our proposed hypothesis for the 
cause of the spectral bluing is described in the following sections.  

\begin{figure*}[t!]
\begin{center}$
\begin{array}{ll}
\includegraphics[width=0.4\linewidth,angle=90]{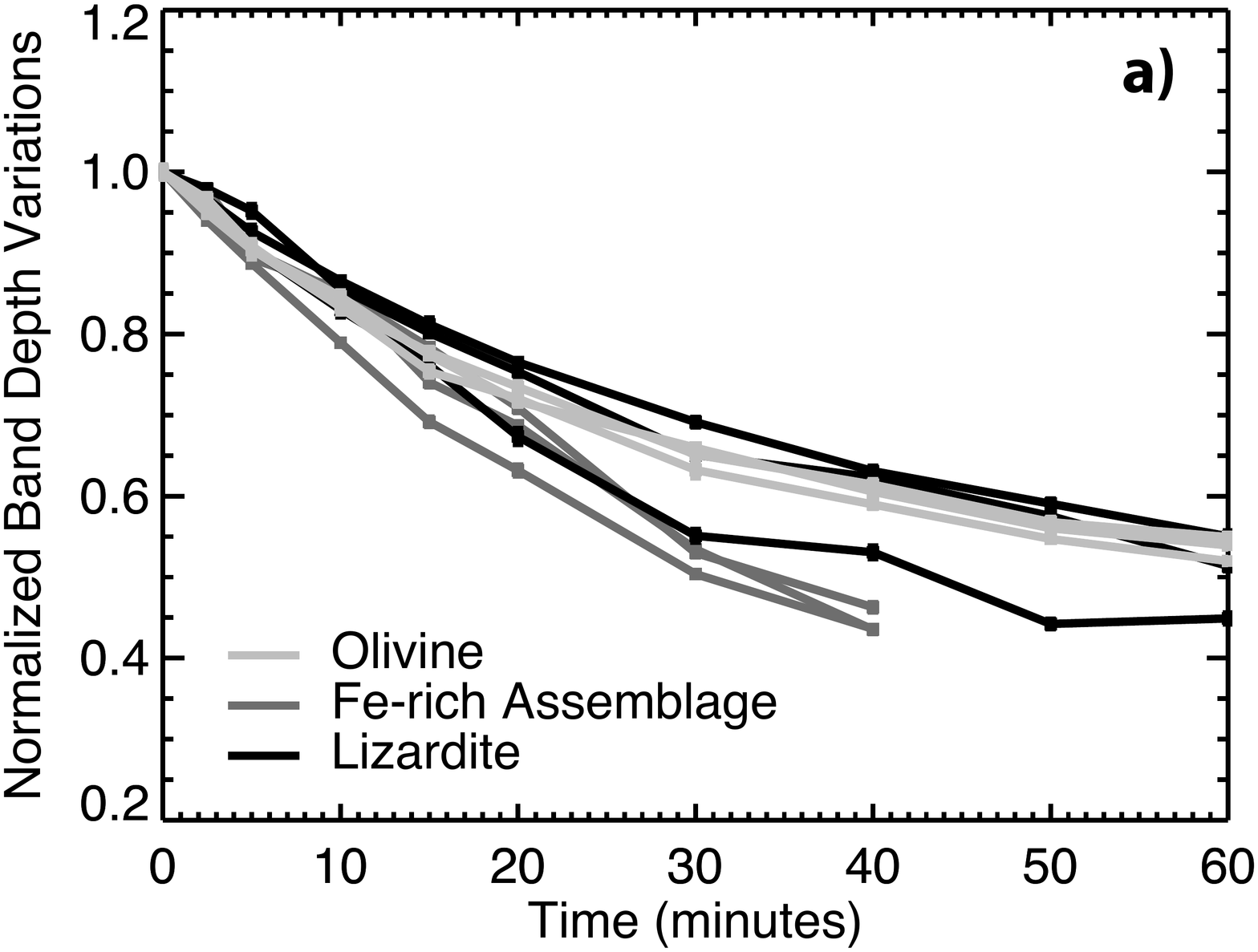} 
\includegraphics[width=0.4\linewidth,angle=90]{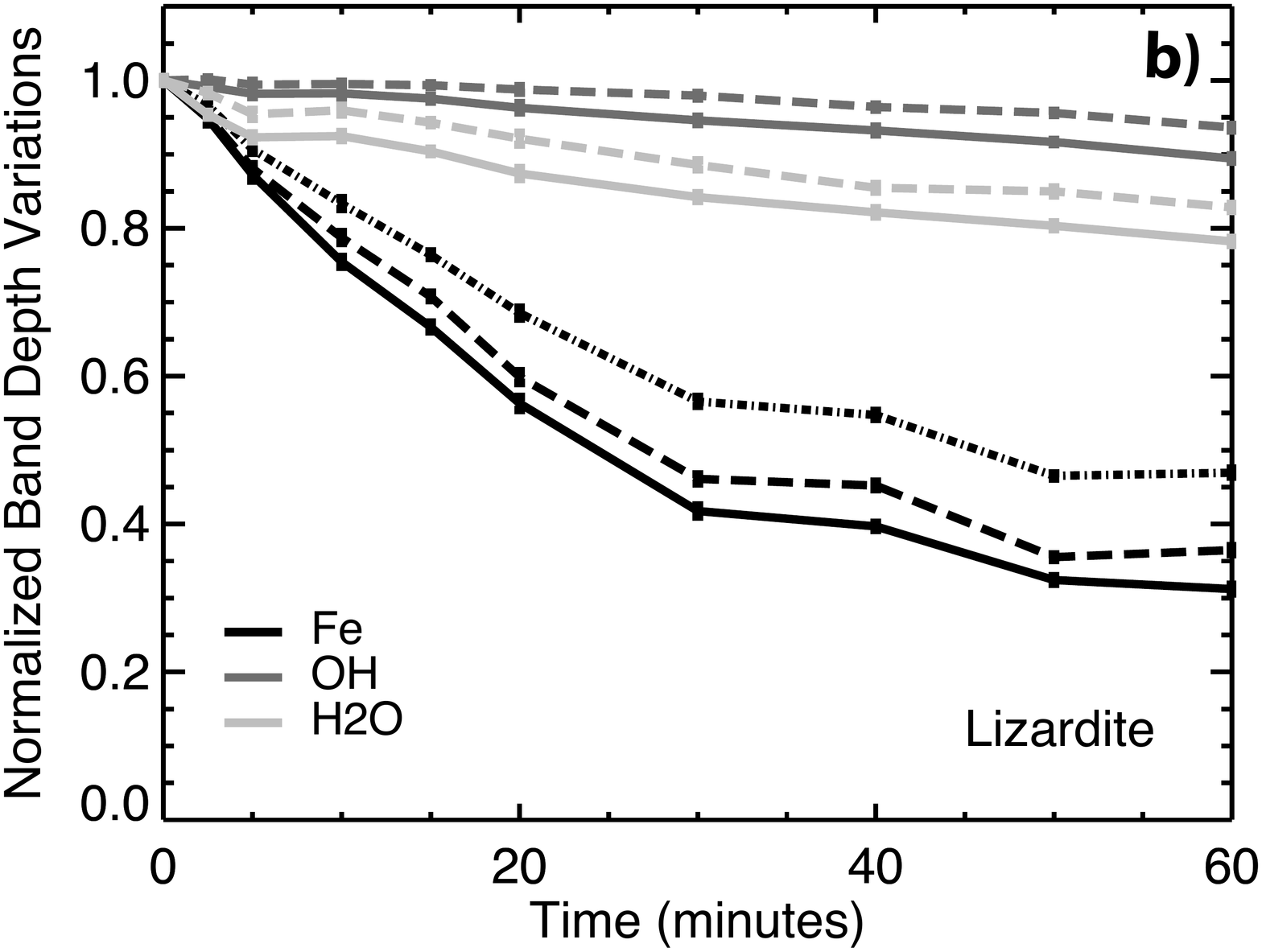} 
\end{array}$
\end{center}
\caption[Band depth variations]{\label{band_variations}
a) The variations in the olivine 1.05 $\mu$m, lizardite 1.13~$\mu$m, and Fe-rich 
assemblage $\sim$2.4~$\mu$m band depths for each of the three experiments.  
The data have been normalized to the initial band depth values of the fresh, non-irradiated samples.  
b) The lizardite band depth variations for 
the Fe$^{2+}\rightarrow$ Fe$^{3+}$ 0.75$\mu$m (solid), 0.92$\mu$m (dashed), 
1.13$\mu$m (dashed-dot), the OH 1.4$\mu$ (solid) and 2.3$\mu$m (dashed) and the 
H$_2$O 1.98$\mu$m (solid) and 2.13$\mu$m (dashed) features.}
\end{figure*}

Band depth measurements for olivine and lizardite 
were made after dividing the full spectrum by a
linear fit to the continuum shoulders on either side of an absorption feature. 
The band depths for the broad Fe$^{2+}$ feature centered at 
1.05 $\mu$m in olivine were derived
using continuum points at 0.70~$\mu$m and 1.60~$\mu$m.  Although the Fe-rich assemblage 
exhibits a broad feature between $\sim$0.7 to 2.0 $\mu$m, 
the lack of clear shoulders and minima 
made characterization in this region difficult.  
Thus, we use the feature near 2.4 $\mu$m to characterize
the band depth changes as a function of irradiation time. However, due to the presence of
only one continuum shoulder, we were limited to using a ratio of the reflectance 
values near the band minimum and shoulder at 2.4 $\mu$m and 2.2 $\mu$m, 
respectively.  Measurements of the lizardite Fe$^{2+}\rightarrow$ Fe$^{3+}$ minima 
at 0.75, 0.92, and 1.13 $\mu$m \citep{Burns:1993} were measured using the 
continuum shoulders at 0.57 and 1.7$\mu$m. 
Figure \ref{band_variations}a shows the relative variations in the olivine 1.05 $\mu$m, 
lizardite 1.13 $\mu$m and the Fe-rich assemblage 2.4 $\mu$m features.  
It is important to note that while band depths variations were observed, 
no significant variations were see in the absorption band center wavelengths. 
Although the slope and albedo variations are quite different for each of
the samples, the Fe-rich assemblage, lizardite and olivine 
all show similar trends in band depth reduction.  

In addition to the Fe$^{2+}\rightarrow$ Fe$^{3+}$ intervalence 
charge transfer (IVCT) features, lizardite
also has a suite of absorption features caused by hydroxyl overtones \citep{Calvin:1997} 
and water \citep{Jain:2012}.  The multiple spectral features provide a useful tool to 
characterize how each of these species varies in response to simulated space weathering. 
Variations in the OH band depths were measured using continuum shoulders 
near 1.35 and 1.45$\mu$ for the 1.4$\mu$ feature, and 2.28 and 2.35$\mu$m 
for the 2.3$\mu$m feature. 
H$_2$O features at 1.98 and 2.13$\mu$m were characterized using continuum shoulders 
near 1.85 and 2.2$\mu$m.  Figure \ref{band_variations}b compares the relative reduction in 
band depths between each of the lizardite IVCT, hydroxyl and water features measured in this study.
The iron features exhibit the greatest band reduction, while the hydroxyl 
features are the least affected by laser irradiation.  
Additionally, the degree of band reduction for all three species decreases at longer wavelengths. 
This wavelength dependency on band reduction is consistent with the production of 
{\it n}pFe$^0$, however it appears that {\it n}pFe$^0$ does not affect all features equally and a compositional
dependency also exists. 

\subsection{Volatile Production}
\begin{figure*}[b!]
\begin{center}$
\begin{array}{ll}
\includegraphics[width=0.39\linewidth,angle=90]{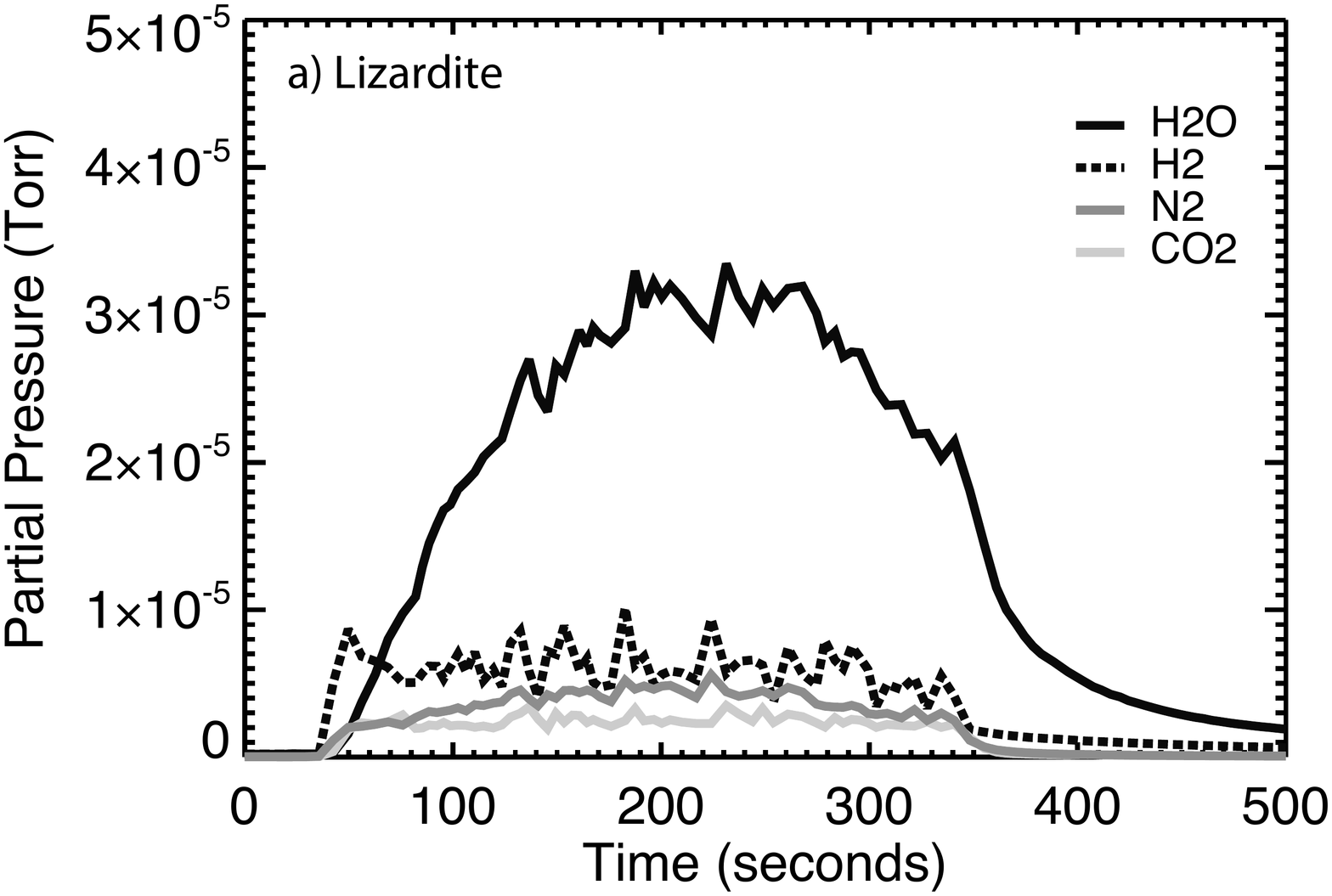}
\includegraphics[width=0.39\linewidth,angle=90]{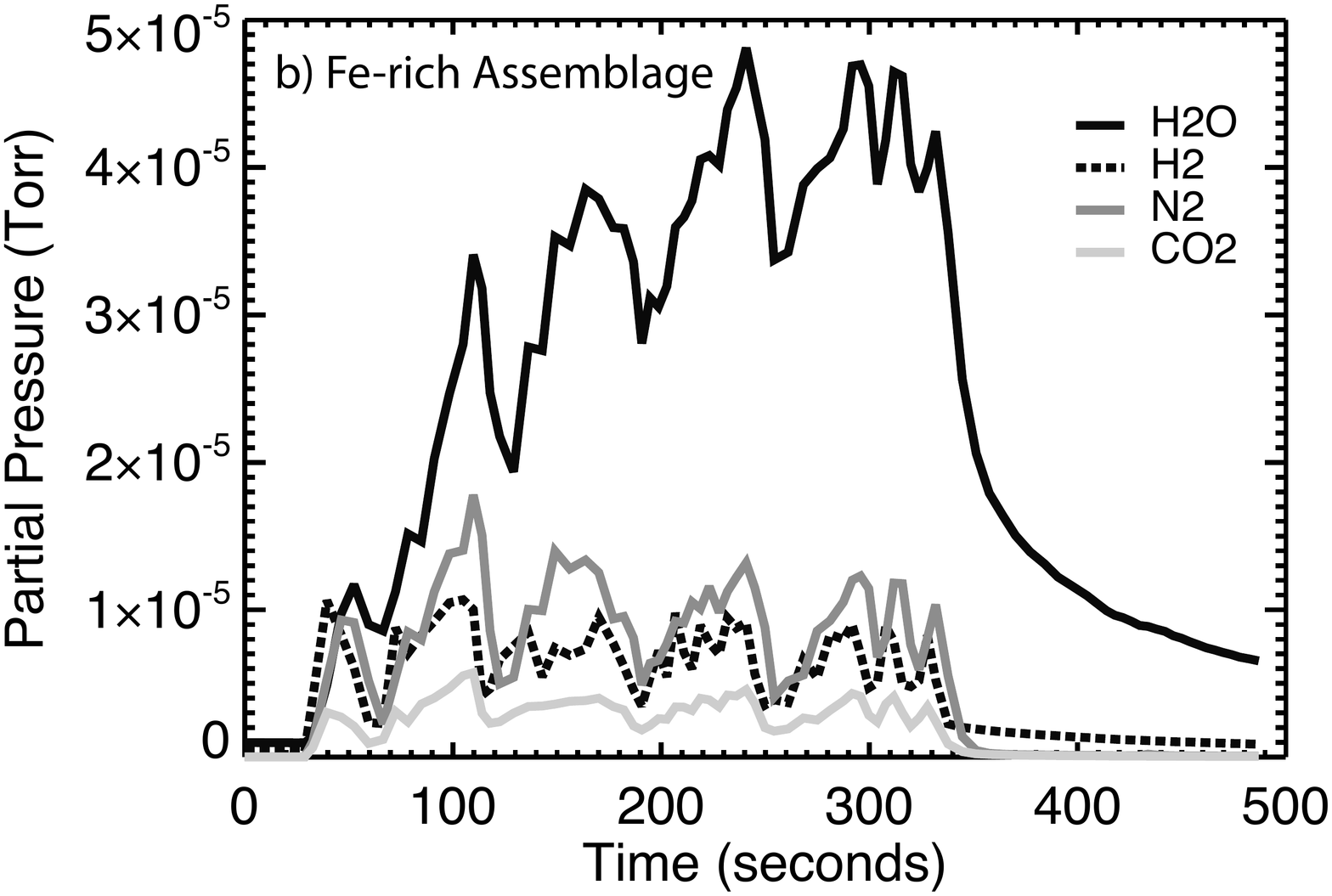} \\
\includegraphics[width=0.39\linewidth,angle=90]{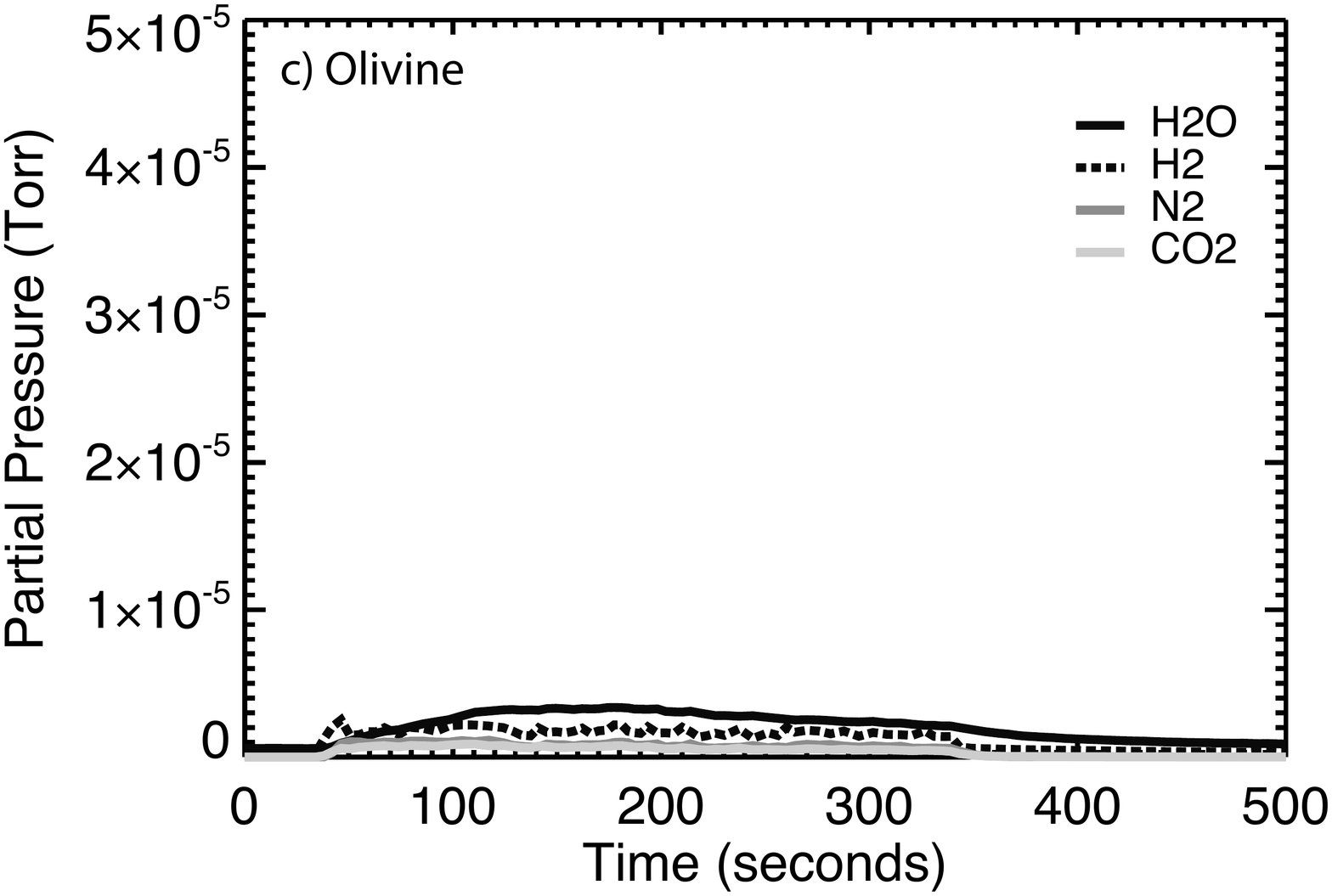}
\includegraphics[width=0.37\linewidth,angle=90]{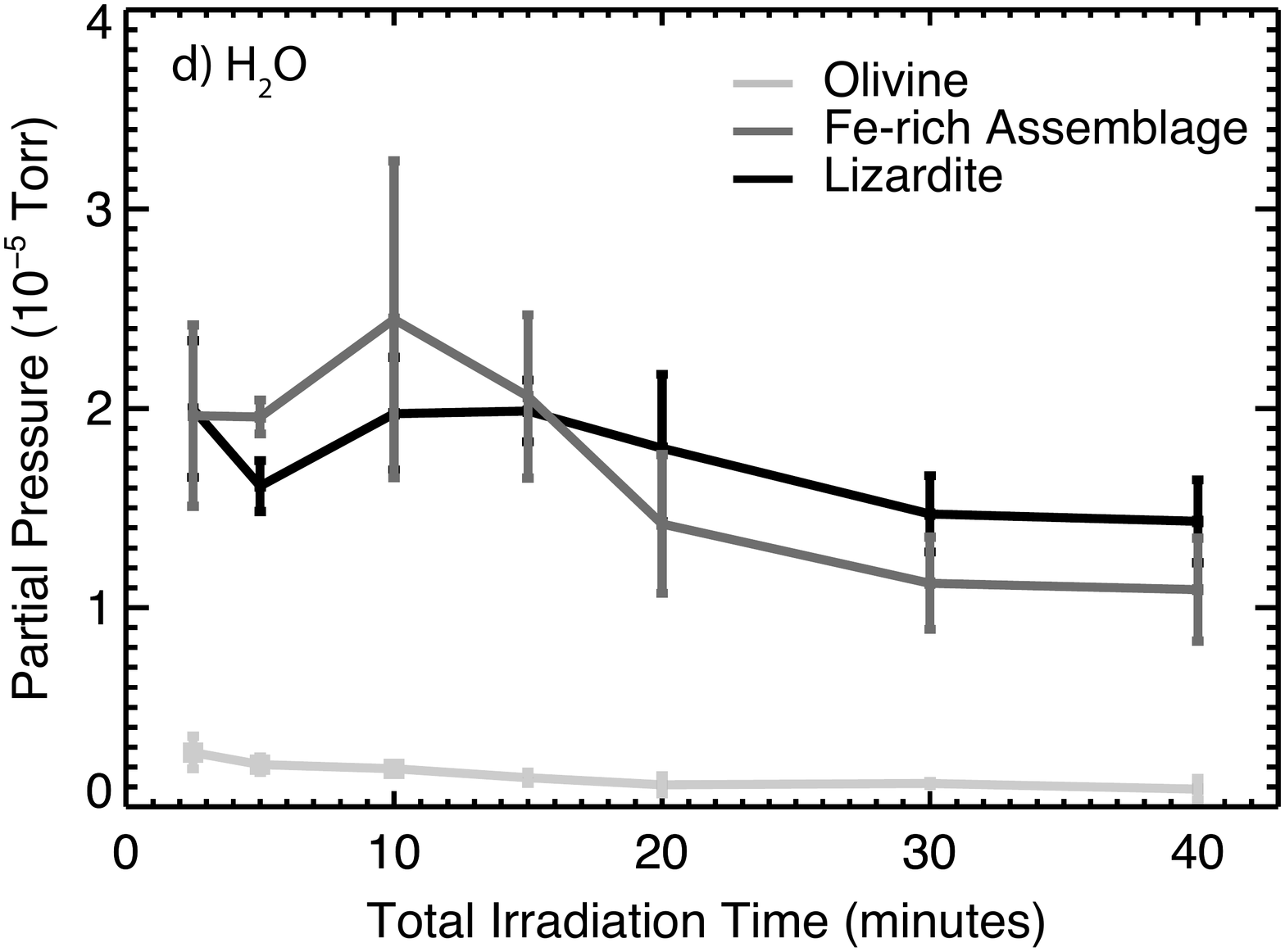} 
\end{array}$
\end{center}
\caption[Emitted gases ]{\label{mass_spec_compare}  
H$_2$O and trapped air constituents released during a 5 minute interval of laser irradiation on 
a) lizardite, b) the Fe-rich assemblage and c) olivine. {\it Note:} The peaks in figure b 
result from the laser spot nearing the 
edge of the sample.  d) The mean partial pressures of 
H$_2$O released by the three minerals for all of the irradiation sets. 
Mean pressures and 1-sigma uncertainties were derived from
the three sets of experiments conducted on each mineral/assemblage.    }
\end{figure*}

Partial pressure measurements for a variety of gases (H$_2$O, N$_2$, CO, CO$_2$) 
released during 5-minute laser irradiation series of olivine, 
lizardite and the Fe-rich assemblage 
are shown in Fig. \ref{mass_spec_compare}a-c.  
When comparing the fluctuations as a function of irradiation 
time, the similarity in fluctuations between the gases and nitrogen suggest the majority of the
species originate from air trapped in the sample powders.  It is also evident that 
the samples appear to trap and release different fractions of air during irradiation, 
suggesting that the `trapping' efficiency is different from mineral to mineral.  

H$_2$O was detected during the irradiation of olivine 
but occurs at levels an order of magnitude less than the other minerals
and likely originates from adsorbed water.  A significant fraction of 
H$_2$O was released from the aqueously altered minerals, which decreases
from the initial to final experiments 
by $\sim$30\% for lizardite and $\sim$50\% for the Fe-rich assemblage.
Reflectance spectra also show a similar degree of reduction in H$_2$O
band depths, with lizardite showing a band reduction of $\sim$20\% 
and the Fe-rich assemblage showing a reduction of $\sim$60\%.  
While there is some consistency 
between the fraction of H$_2$O released and the band depth variations, 
it is important to note that there is also a wide spread in the partial 
pressure values across the three experiments for each irradiation 
sequence (Fig. \ref{mass_spec_compare}d).
These large standard deviations likely arise from variations in the fraction of water 
that is re-adsorbed by the samples when they are taken out of the vacuum 
chamber for spectral imaging, however quantifying the fraction of the 
native versus adsorbed water lost during the 
irradiation is beyond the scope of this study.  Nevertheless, in contrast to previous 
interpretations \cite[e.g.][]{Rivkin:2002}, these data 
show that the liberation of water and the creation of {\it n}pFe$^0$ and other 
space weathering products can occur in conjunction.
In other words, the samples do not have to be dehydrated before the creation 
of {\it n}pFe$^0$ can occur. 

\subsection{SEM and TEM sample characterization}
\begin{figure*}[h!]
\begin{center}$
\begin{array}{ll}
\includegraphics[width=0.5\linewidth,angle=0]{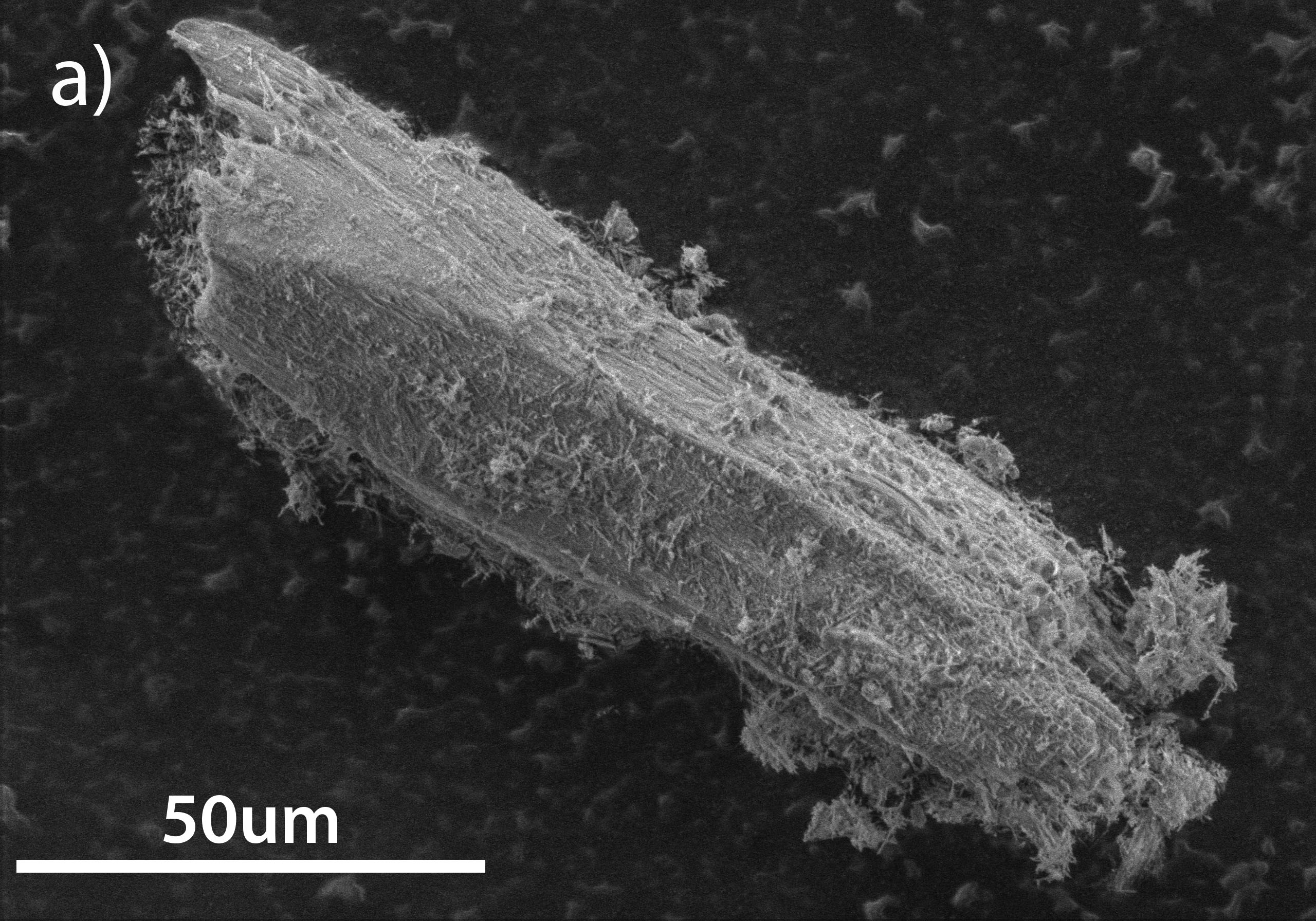}
\includegraphics[width=0.5\linewidth,angle=0]{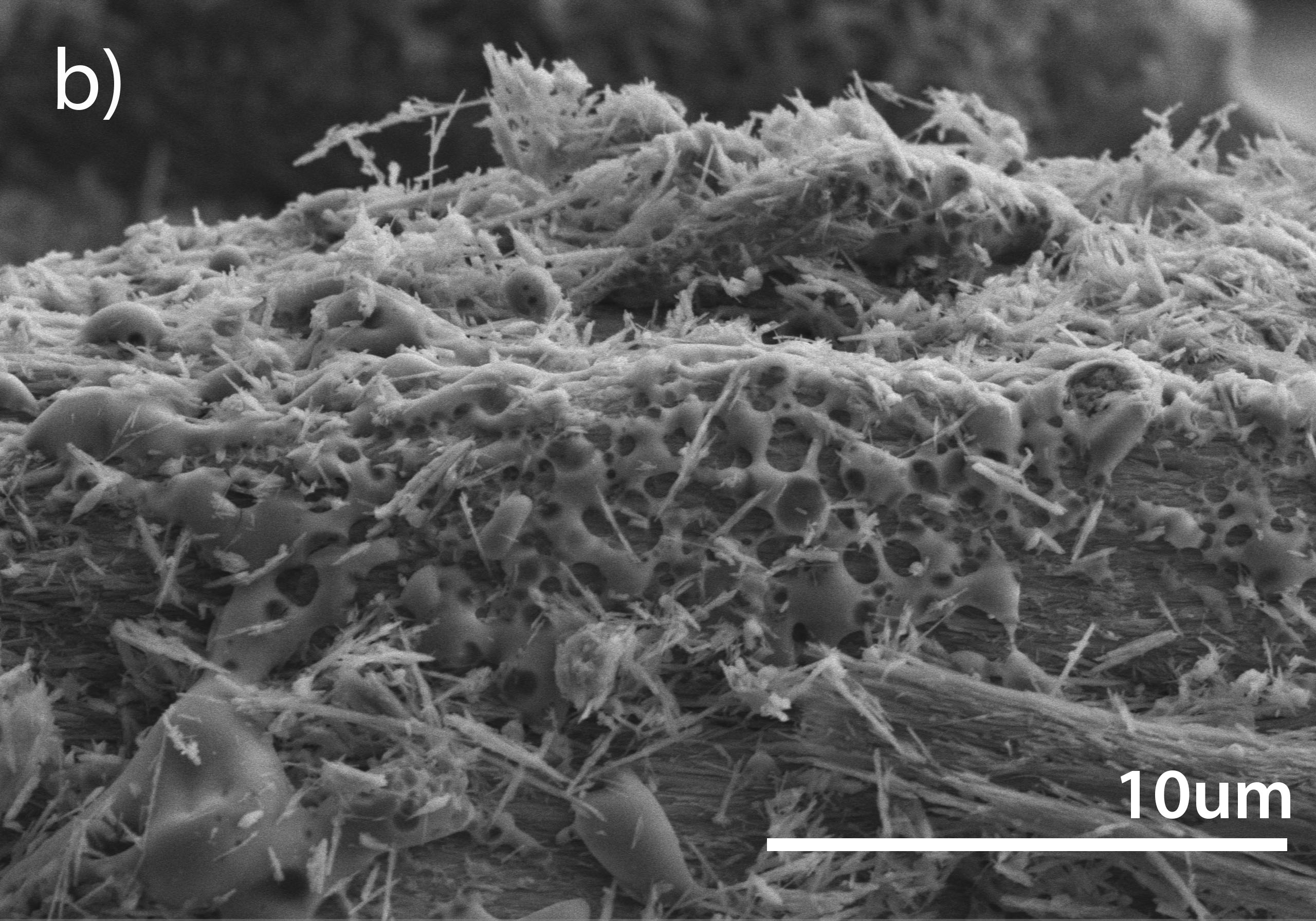} \\
\includegraphics[width=0.5\linewidth,angle=0]{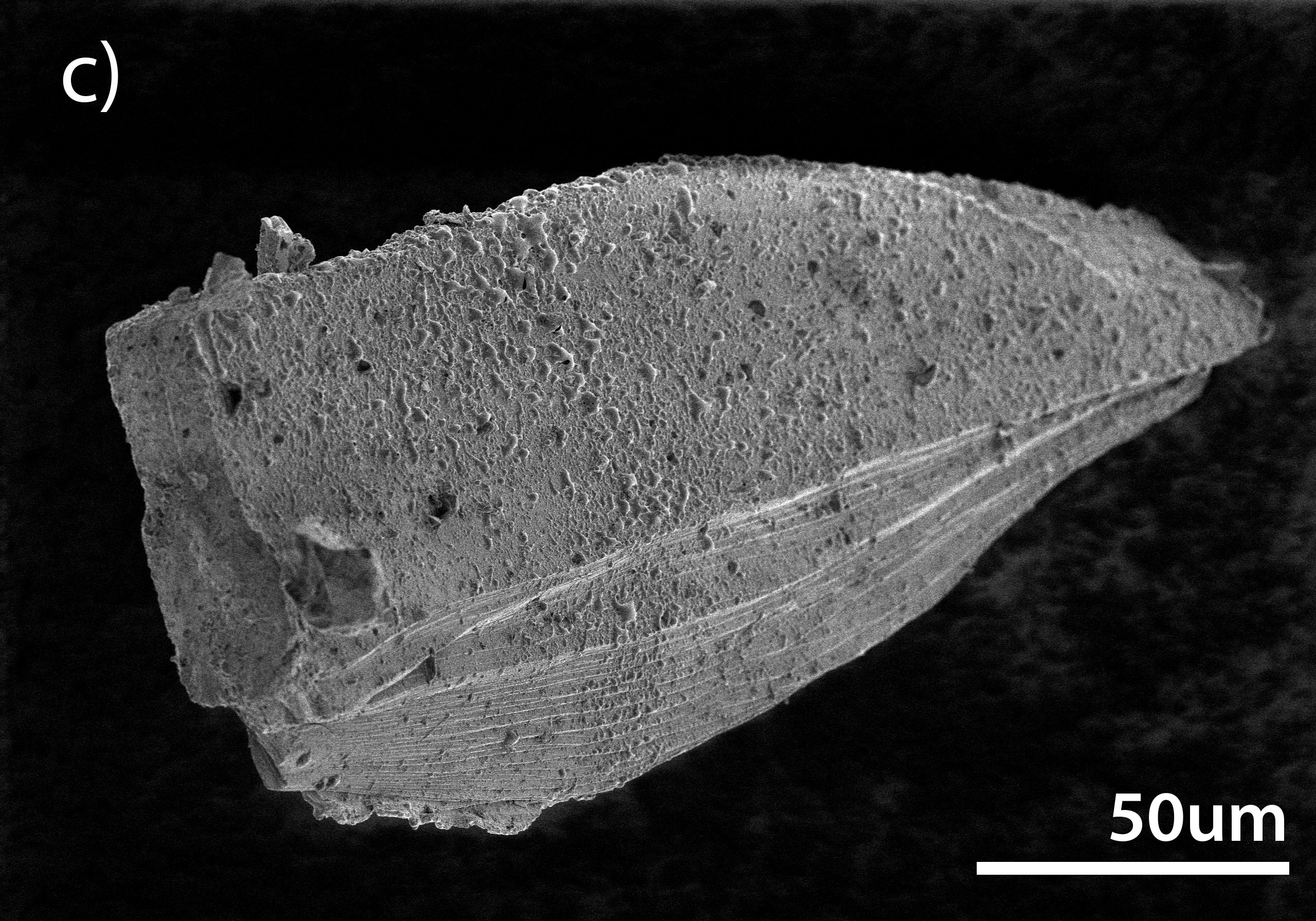}
\includegraphics[width=0.5\linewidth,angle=0]{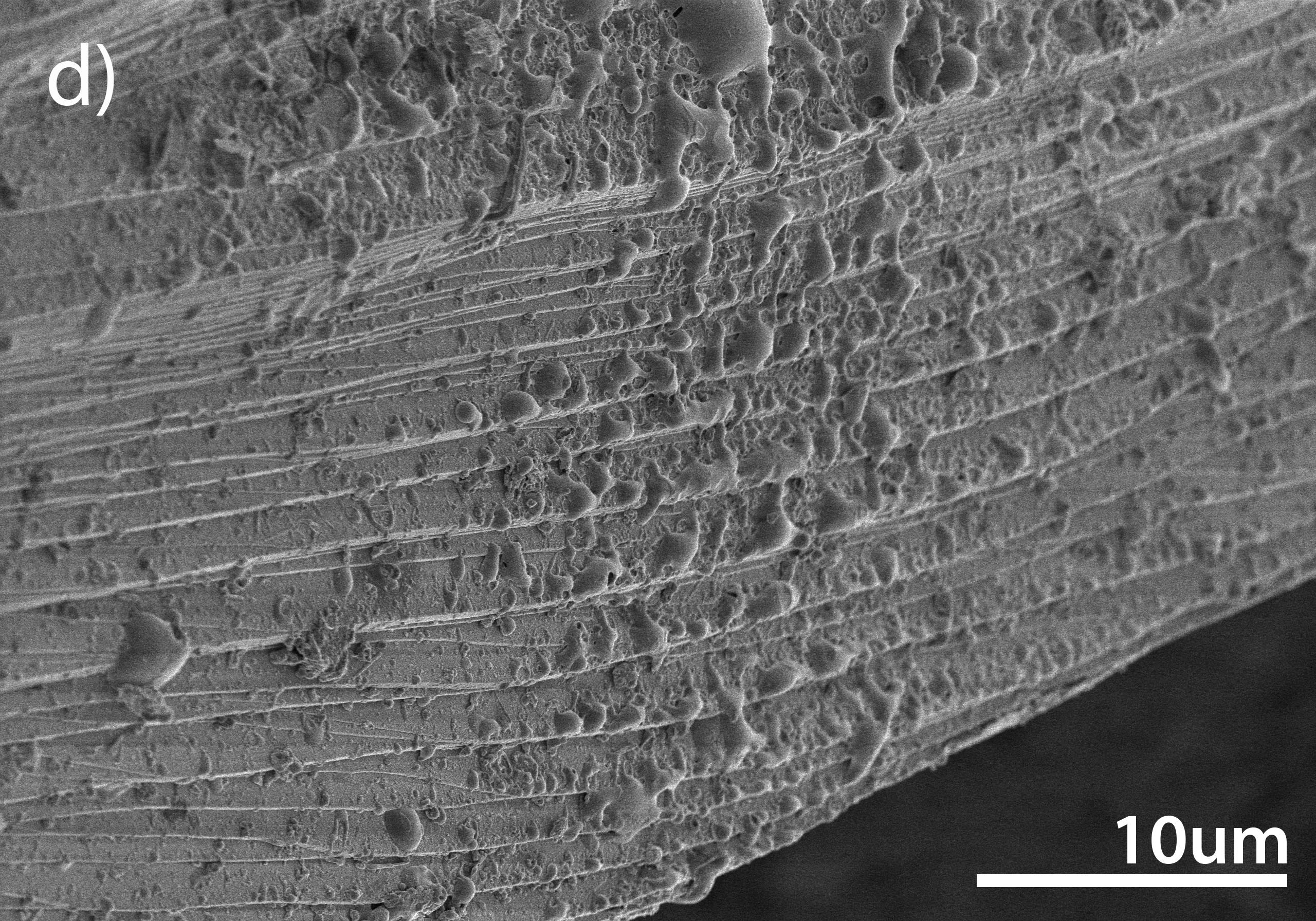} 
\end{array}$
\end{center}
\caption[SEM images ]{\label{sem_images}  
Secondary electron images of laser-irradiated lizardite and 
Fe-rich assemblage grains taken with the PBRC SEM.  Images of 60 minute irradiated lizardite 
whole grain (a) and grain surface features (b).  Images of a cronstedtite grain (c) 
and grain surface features (d) from a 40-minute irradiated Fe-rich assemblage sample.
}		
\end{figure*}

SE images of the 60 minute irradiated lizardite 
and 40 minute irradiated Fe-rich assemblage grains both show widespread 
melt deposition, however, there is a distinct difference in the physical appearance of the 
two samples (Fig. \ref{sem_images}a-d).  
Particularly noticeable in Fig. \ref{sem_images}a and 
\ref{sem_images}b are very fine 
lizardite needles, only a few microns in length, 
coating the melt deposition regions on the larger host grain.  Due to the 
fibrous nature of lizardite, these needles are
likely remaining pieces of the irradiated portion of the sample that redeposited on nearby
grains.  These fine fibers appear to be covering a significant fraction of the
melt deposits and may be
impeding the optical effects of {\it n}pFe$^0$ on the spectral features of
the lizardite samples. In particular,
the contrasting behavior between the significant reduction in
the Fe$^{2+}\rightarrow$ Fe$^{3+}$ IVCT band depths and the 
minimal changes in albedo and slope may be a result of the fibers 
preferentially reducing some optical properties of {\it n}pFe$^0$ over others. 

\begin{figure*}[t!]
\begin{center}$
\begin{array}{llllll}
\includegraphics[width=0.5\textwidth,angle=0]{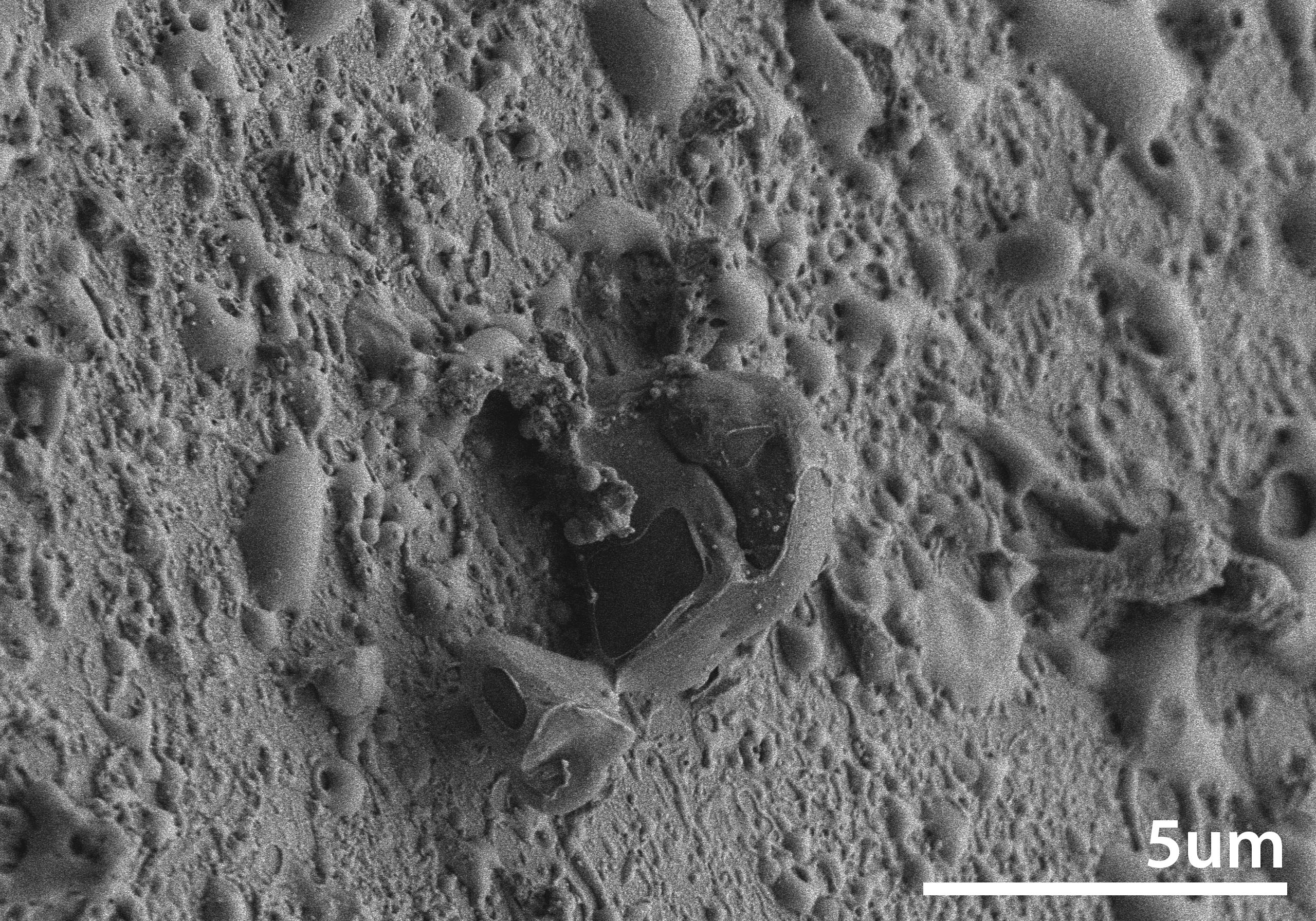} 
\includegraphics[width=0.5\textwidth,angle=0]{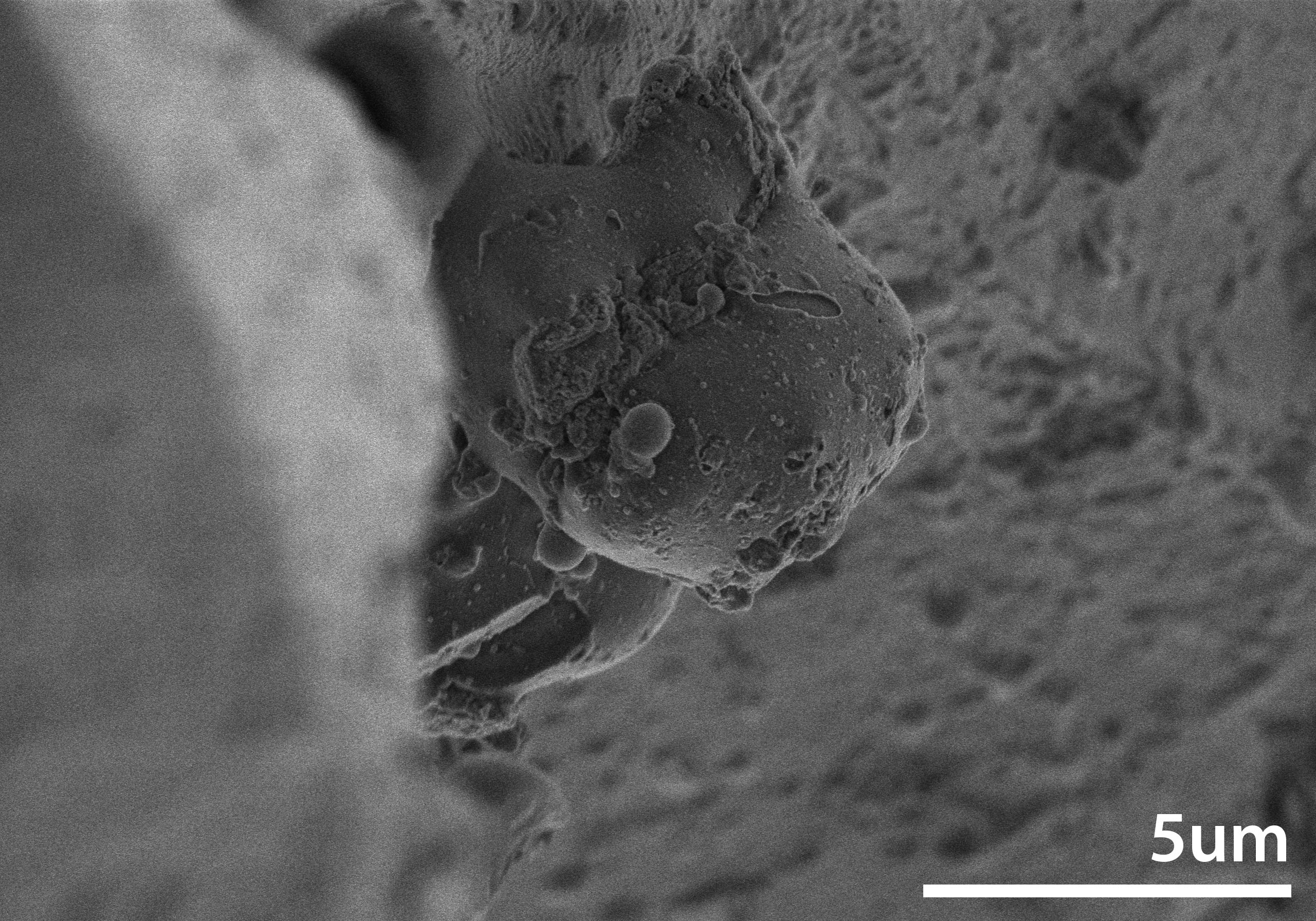}\\
\end{array}$
\end{center}
\caption[Element maps of weathered cronstedtite]{\label{sem_fibs}  
SE images of the two dark grains from the irradiated 
Fe-rich assemblage that were cut into FIB sections and analyzed with 
(S)TEM.}
\end{figure*}

\begin{figure*}[t!]
\begin{center}$
\begin{array}{llllll}
\includegraphics[width=0.5\textwidth,angle=0]{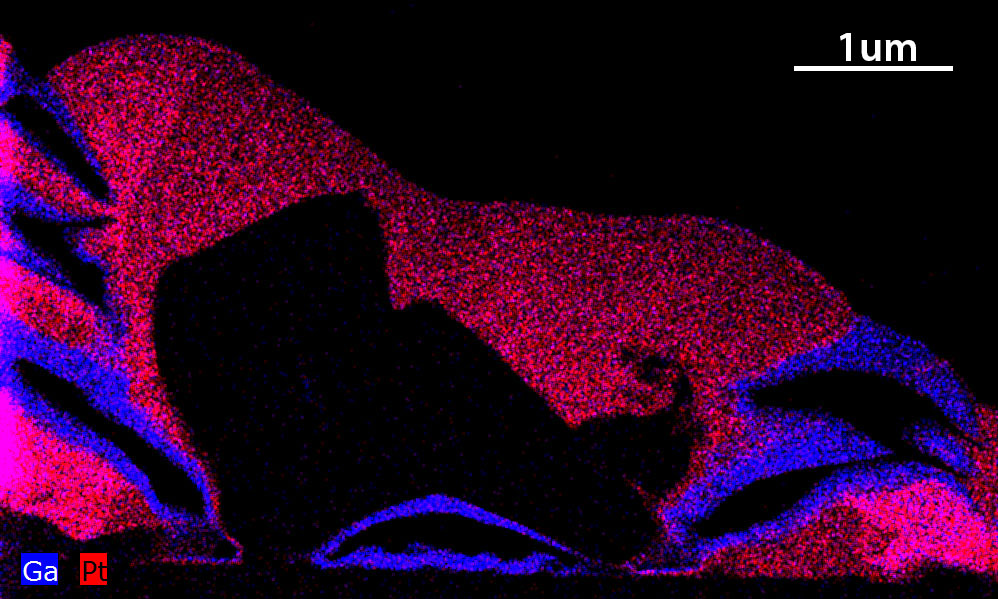} 
\includegraphics[width=0.5\textwidth,angle=0]{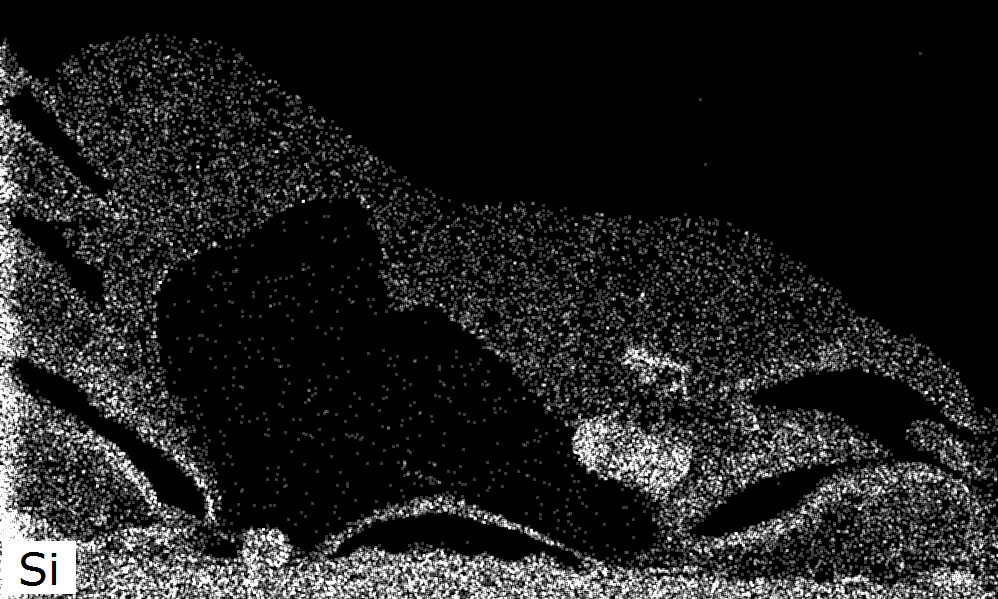}\\
\includegraphics[width=0.5\textwidth,angle=0]{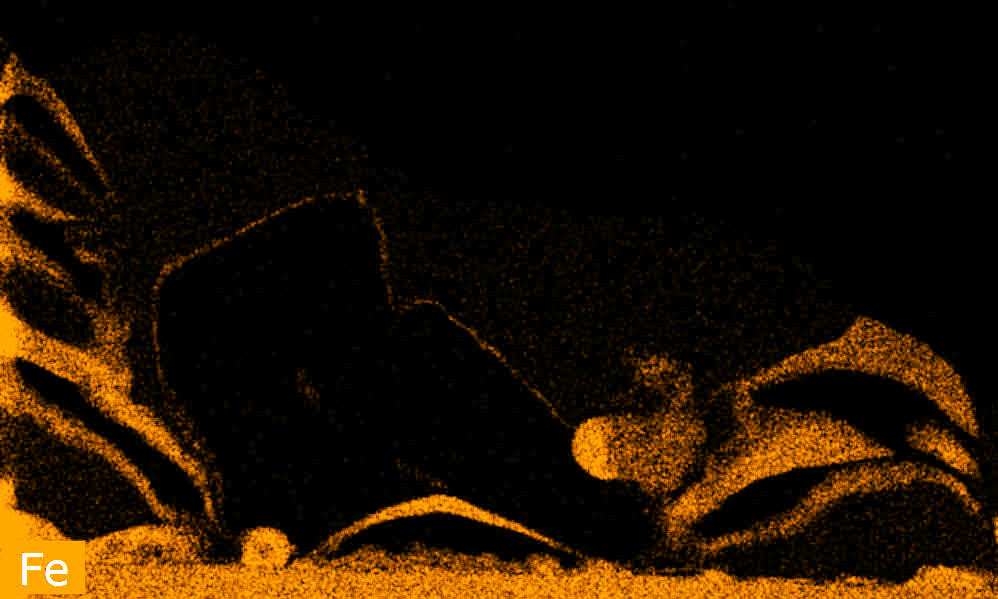} 
\includegraphics[width=0.5\textwidth,angle=0]{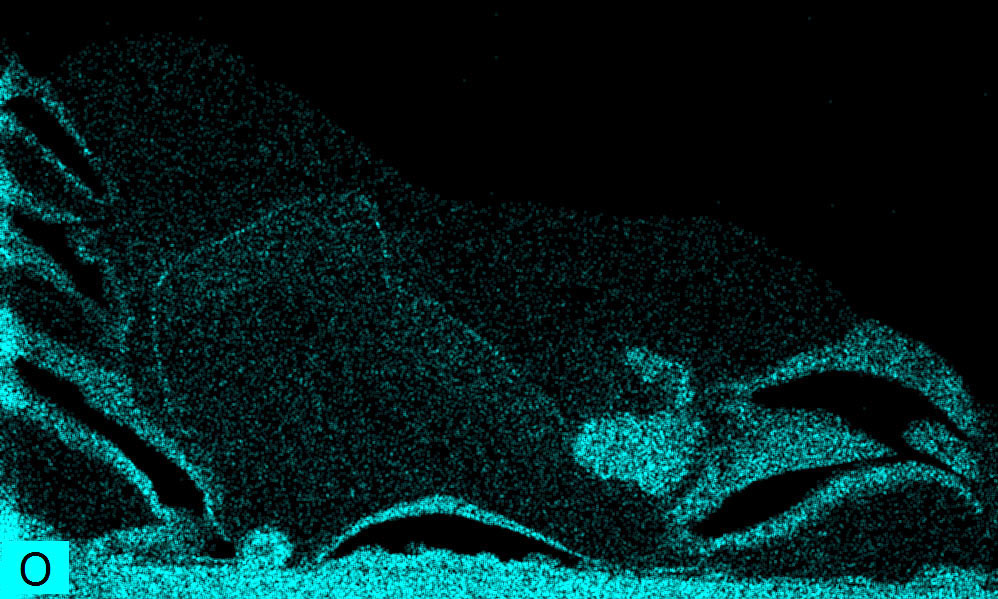} \\
\includegraphics[width=0.5\textwidth,angle=0]{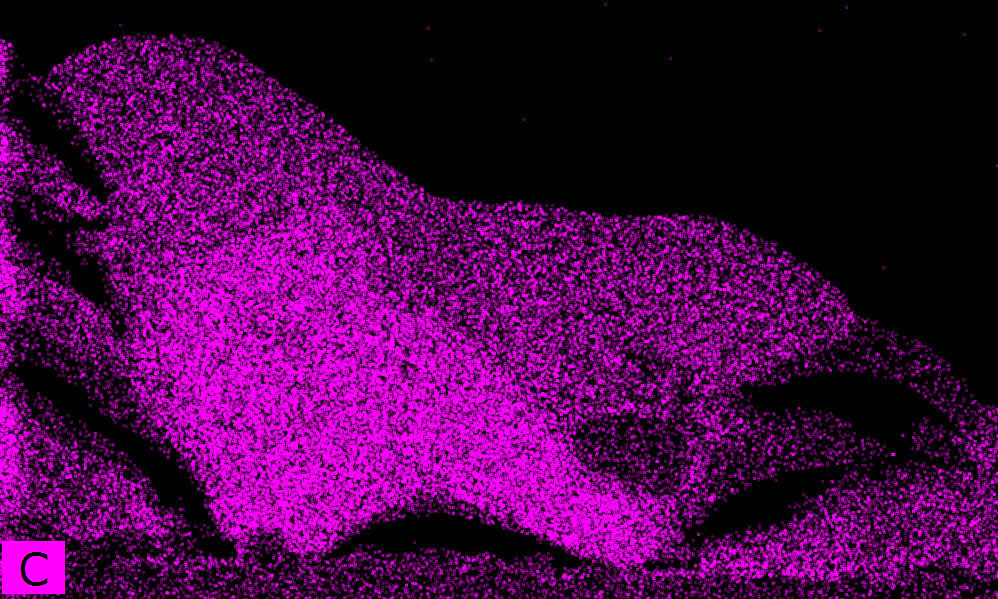} 
\includegraphics[width=0.5\textwidth,angle=0]{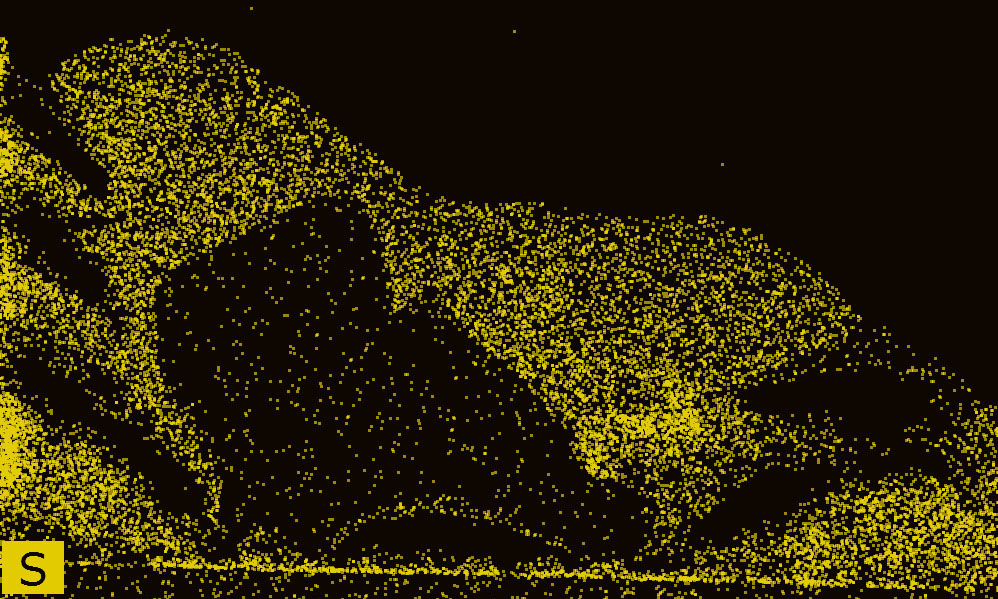}  \\
\end{array}$
\end{center}
\caption[Element maps of weathered cronstedtite]{\label{element_map1}  
OSU Titan element maps of FIB section 1 created from an Fe-rich 
assemblage melt product.  The Ga/Pt map references regions where
materials used to create the FIB section are located and thus 
should be disregarded in other images. Analyses of the C-rich region
shows the presence of C, N and O. }
\end{figure*}

Figures \ref{sem_images}d and \ref{sem_fibs} show 
two characteristic types of deposition materials
observed through the Fe-rich samples.  The first type
(Fig. \ref{sem_images}d) resemble the melt deposits seen on 
the lizardite grains, whereas the second type (Fig.  \ref{sem_fibs}) 
consists of larger, micron sized dark particles.  
Element maps of the first FIB section (FIB section 1)
taken with the OSU Titan (Fig. \ref{element_map1}) 
show that the dark grain is a large carbon-rich particle coated by a thin 
layer of Fe, O and Si.  The Pt and Ga map
represents regions to ignore as this is where sputtering during the 
ion milling process leads to the redeposition of Pt, Ga and bulk sample atoms 
in void spaces. 
Adjacent to the C grain are two small, 
Fe rich melt droplets that contain Si, O, and S (Table \ref{tab_edx}).  Unexpectedly, 
EDX analysis of the C-rich region reveal the presence of nitrogen.  
To confirm the nitrogen signature, EDX analysis was
repeated on the C-grain using the superior NCEM TitanX.  Both analyses are
given in Table \ref{tab_edx}, and while the first analysis 
detected small amounts of Si and Fe, it is likely from the redeposition of 
FIB sputtered material in a narrow void region.  The second analysis shows that the 
grain is purely composed of C, N and O when the void region is excluded.  
These results suggest the grain may
be some form of an organic compound or a synthesized heterofullerene.

\begin{figure*}[t!]
\begin{center}$
\begin{array}{llllll}
\includegraphics[width=0.45\textwidth,angle=0]{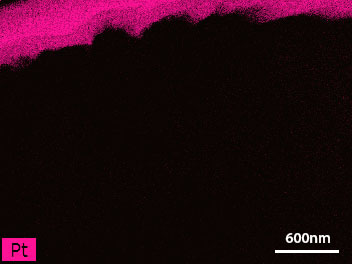} 
\includegraphics[width=0.45\textwidth,angle=0]{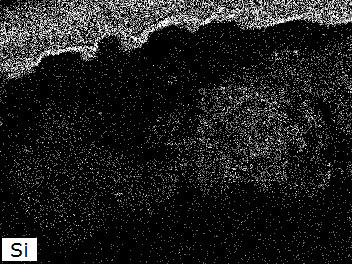}\\
\includegraphics[width=0.45\textwidth,angle=0]{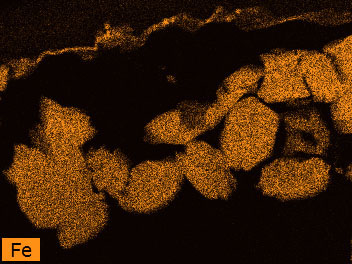} 
\includegraphics[width=0.45\textwidth,angle=0]{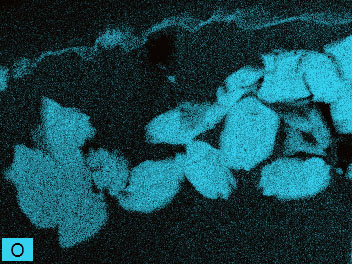} \\
\includegraphics[width=0.45\textwidth,angle=0]{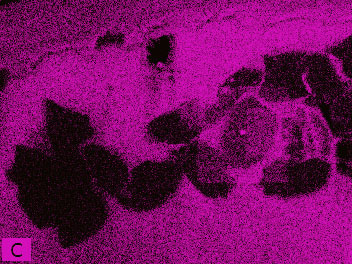} 
\includegraphics[width=0.45\textwidth,angle=0]{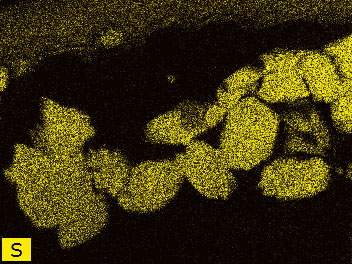}  \\
\end{array}$
\end{center}
\caption[Element maps of weathered cronstedtite]{\label{element_map2}  
NCEM TitanX EDX maps of FIB section 2 created from an Fe-rich 
assemblage melt product.  Oxidized pyrite grains are embedded in a carbon-rich 
melt grain.  The Pt map references regions where
materials used to create the FIB section are located and thus 
should be disregarded in the elemental maps. }
\end{figure*}

Elemental maps of the second FIB section (FIB section 2) 
are shown in Figure \ref{element_map2}. 
To assess whether the nitrogen content in the FIB section 1 was anomalous, 
EDX analysis was performed on FIB section 2 with the NCEM TitanX.
Again nitrogen was found in the C-rich region of the grain (Tab. \ref{tab_edx}), 
suggesting that N may be common among these dark particles.  
The semi-continuous glassy rim surrounding the 
C-rich grain was found to contain oxidized rather than metallic Fe.  
The Fe particles were likely oxidized during irradiation, 
however there remains the possibility that oxidation occurred when 
the samples were removed from the vacuum chamber, or after sample processing by FIB. 
Analyses of a relatively large ($\sim$50nm) droplet in the 
rim was found to contain C, N, O, Fe, and S.  
The presence of N in the melt rim of the FIB section 2 (Tab. \ref{tab_edx}) 
as well as the carbon grains 
show that the nitrogen content is not anomalous and likely common  
throughout the laser-irradiated samples.  Data 
collected with the RGA show the release of atmospheric species (e.g. N and O) 
from our samples during irradiation, thus the nitrogen in these grains likely comes from reactions 
with the local atmosphere created during irradiation. 

\subsection{Radiative Transfer Model Abundances}
\begin{figure*}[ht!]
\begin{center}$
\begin{array}{ccc}
\includegraphics[width=0.4\linewidth,angle=90]{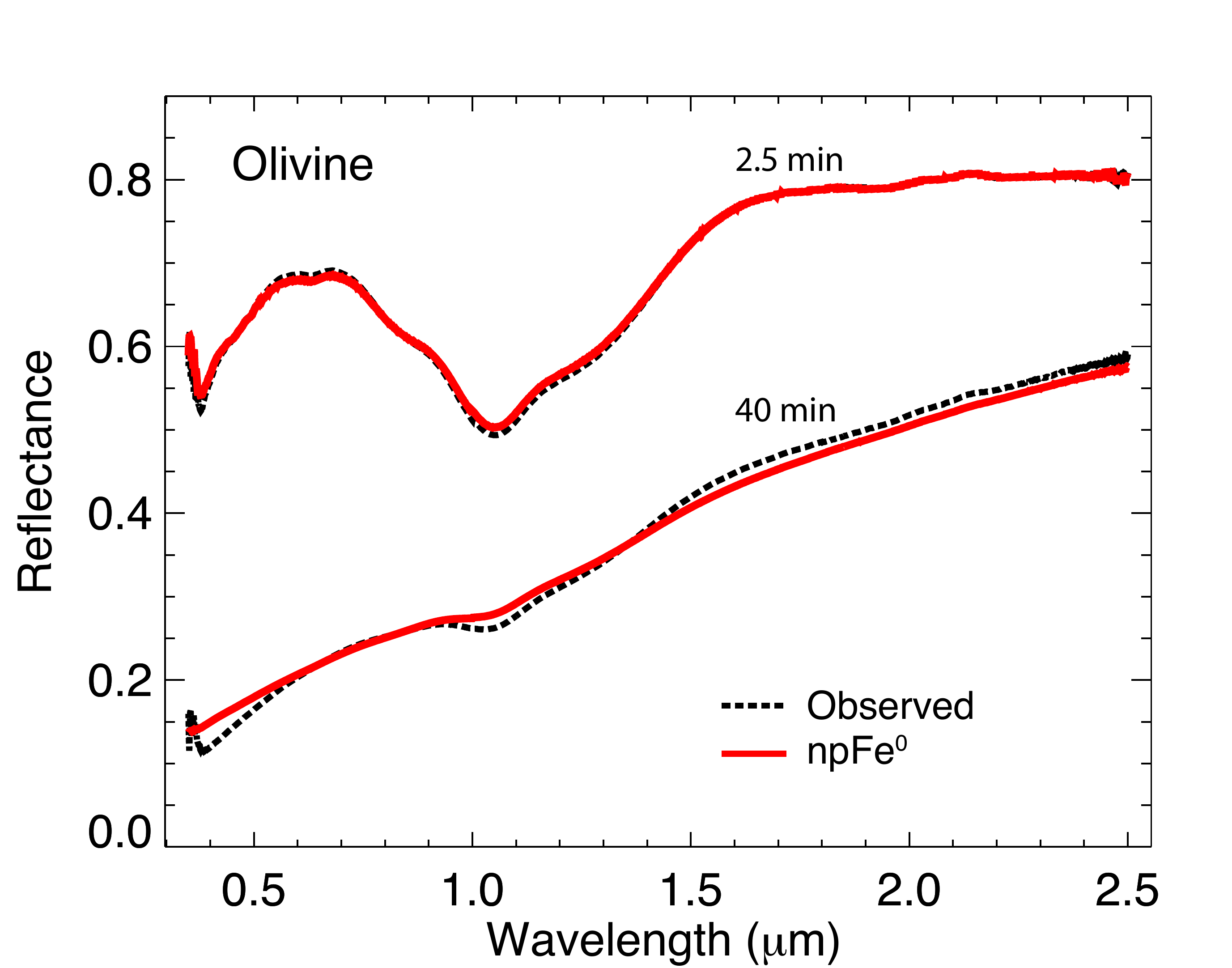} 
\includegraphics[width=0.4\linewidth,angle=90]{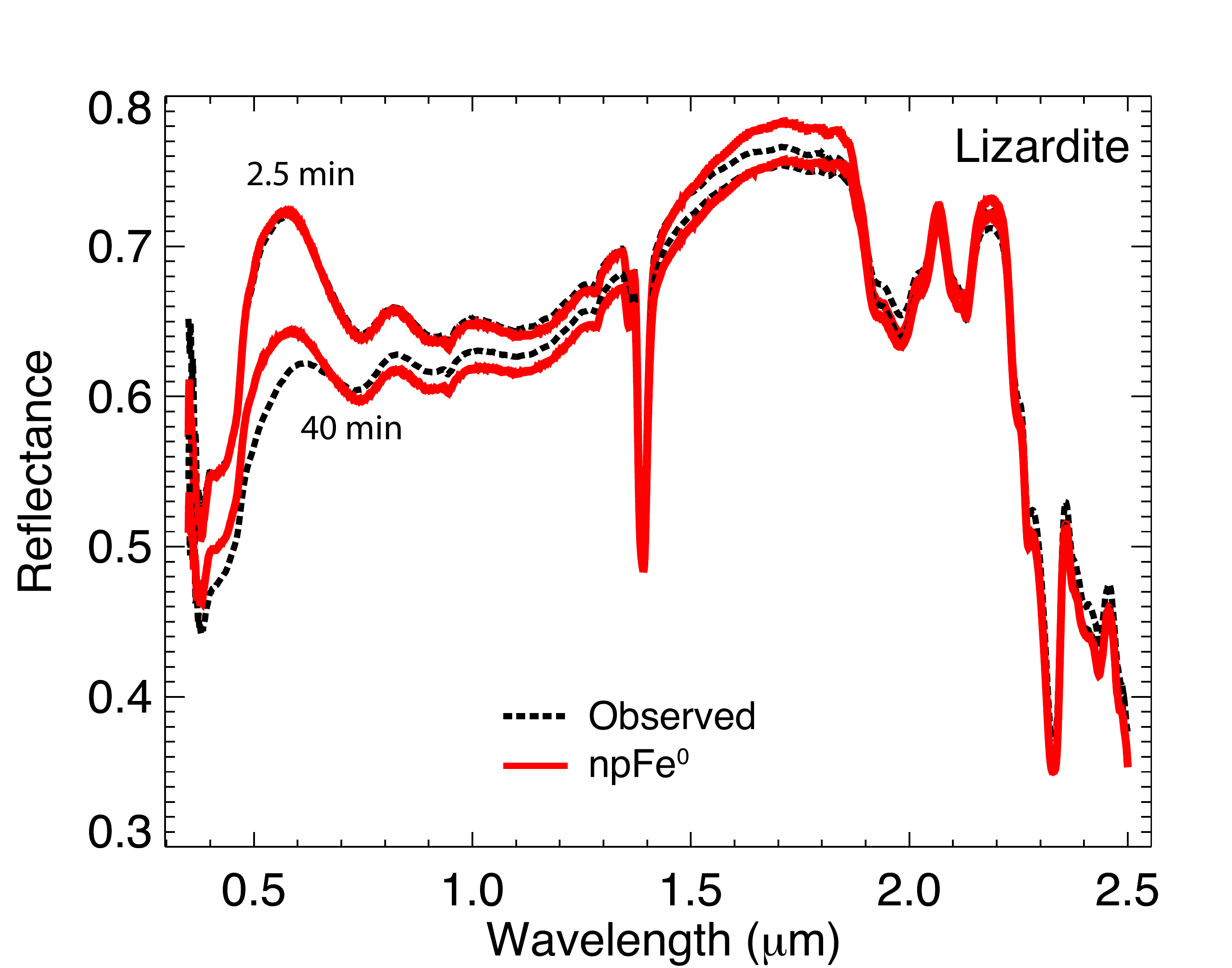} \\
\includegraphics[width=0.45\linewidth,angle=90]{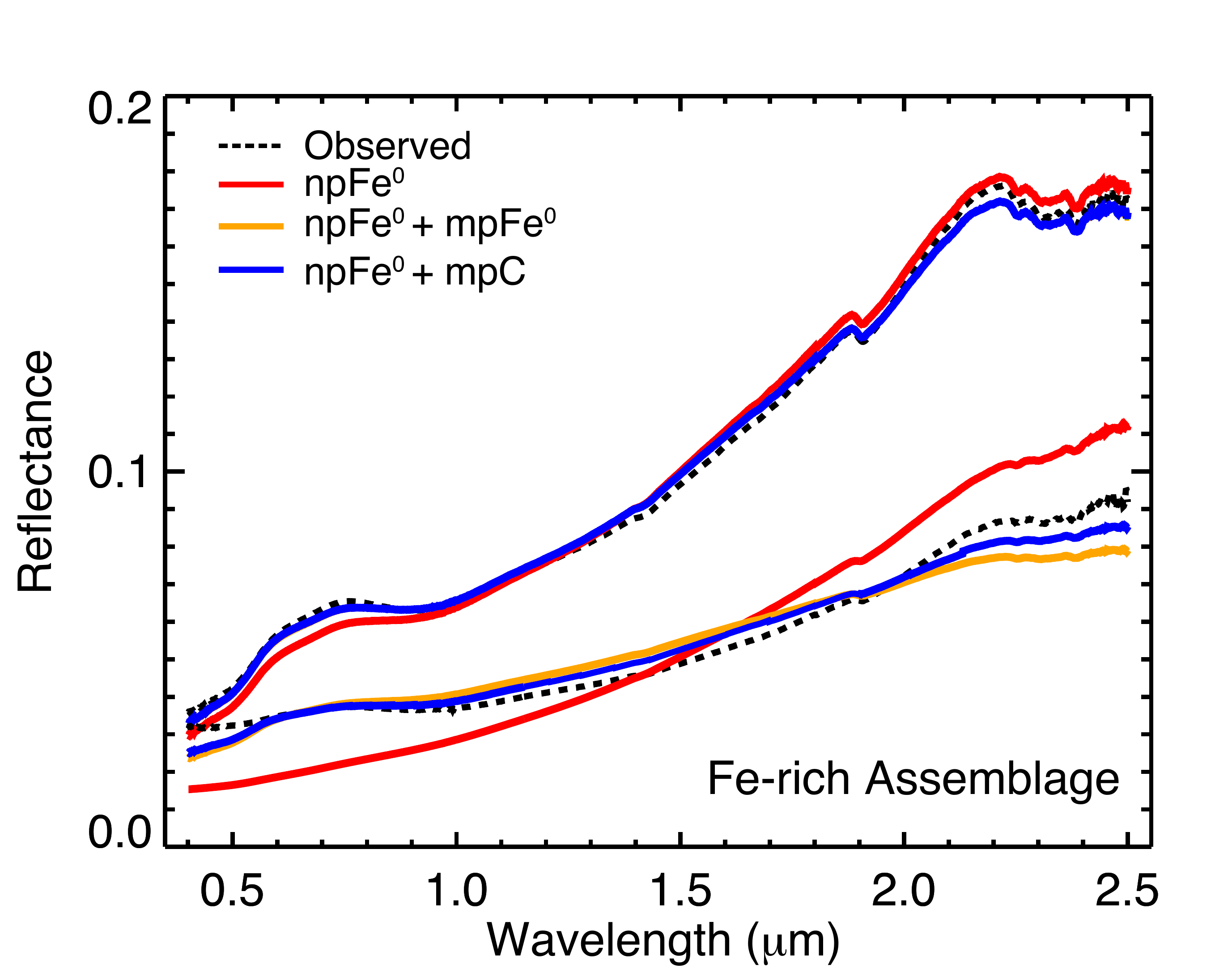} 
\end{array}$
\end{center}
\caption{\label{sw_models1}  
Observed olivine, lizardite and Fe-rich assemblage reflectance spectra and best fit 
radiative transfer models produced using the abundances given in 
Table \ref{tab_model_abundances}. For clarity, only the 2.5 and 40 minute 
(total irradiation time) spectra and corresponding models are shown. {\it Note: 
the 2.5 {\it n}pFe$^0$ and $\mu$pC model (blue) overlaps the 
{\it n}pFe$^0$ and  $\mu$pFe$^0$ model (orange). } }
\end{figure*}

The modified Hapke modeling allows us to study not only 
the optical effects of the nanophase and microphase particles, but 
also the evolution of and relationship between these
particles.  Figure \ref{sw_models1} shows the best fit radiative transfer models for  
the initial and final irradiated spectra for each sample. 
Although both {\it n}pFe$^0$ and $\mu$pFe$^0$ were considered in the model inputs, the  
addition of $\mu$pFe$^0$ did not improve the fits for the olivine and lizardite spectra. 
The best fit model abundances for each sample spectrum as a function of 
irradiation time are given in Table \ref{tab_model_abundances}. 
As is evident from Fig. \ref{sw_models1}b, the model does a poor job of 
reproducing the 40 minute (total) irradiated lizardite spectrum.  The SEM 
images of lizardite (Fig. \ref{sem_images}) revealed that the melt deposits  
were partially coated in very fine lizardite needles.  The poor fits of our model may 
result from the potential effect of these needles on the effects of {\it n}pFe$^0$ or 
the limitations of our model, which assumes spherical host grains and nanoparticles. 
Thus, the {\it n}pFe$^0$ abundances in Table \ref{tab_model_abundances} 
are simply used as broad approximations for the production of {\it n}pFe$^0$ during 
irradiation. 

Due to the presence of micron sized carbon particles,
the Fe-rich assemblage models consider the optical effects of 
both iron and carbon particles (Fig. \ref{sw_models1}c).  We find that 
models using only {\it n}pFe$^0$ 
particles reproduce the spectral reddening reasonably well, but  
do a poor job of reproducing the spectral variations once bluing 
occurs.  The addition of $\mu$pFe$^0$ improves
the model fits, however we find the best spectral fits are obtained 
using a combination of {\it n}pFe$^0$ and $\mu$pC particles.  
Some {\it n}pC particles were observed 
in the rim of one dark melt grain, however these particles 
do not improve the model fits, which may suggest that the carbon particles are typically 
larger, micron sized particles.  
Our best fit models show that {\it n}pFe$^0$ is the dominant particle 
produced during the initial set of irradiation sequences and corresponds to the 
initial spectral reddening of the samples. 
The models also show that $\mu$pC particles quickly become more 
abundant beyond $\sim$15 minutes of irradiation, which coincides with the observed
spectral bluing.  Thus, our models suggest that the {\it n}pFe$^0$ particles 
produce the observed spectral reddening 
whereas the $\mu$pC particles are largely responsible for 
the spectral bluing in our Fe-rich samples. 

It is important to note that while the best fit models 
show an increase in the fraction of {\it n}pFe$^0$ and $\mu$pC with increasing irradiation time, 
the abundances of {\it n}pFe$^0$ begins to decrease beyond $\sim$15 minutes and 
reduces to zero for the final irradiated spectrum
(Table \ref{tab_model_abundances}).  This peculiar trend may result
from the $\mu$pC particles dominating the optical properties of the sample 
once a critical abundance is reached, but it may also be a product of 
using Hapke's model, which was designed for 
use with transparent materials, on opaque minerals.  
However, while our results may be further improved by using a 
more complex radiative transfer model, the models presented here 
are able to reproduce the spectral variations of our 
Fe-rich samples reasonably well.  So although the modeled {\it n}pFe$^0$ and 
$\mu$pC abundances may not reflect the true abundances in our samples, 
our results still provide values insights into the 
relative role that these particles play in the spectral behavior of our samples. 

\section{Discussion}
\subsection{Abiotic organic synthesis resulting from space weathering}
Although airless bodies are exposed to the harsh environment of space, 
impacts, cosmic radiation and other space weathering processes
supply a source of energy that can be used to drive chemical reactions on 
the surfaces of these bodies.  Of particular interest is 
the synthesis of organic compounds on extraterrestrial bodies as they 
may play an important role in the delivery of prebiotic material 
to the early Earth \citep{Chyba:1992}. 
Initial experiments simulating the evolution of
organic molecules used ion bombardment of carbon-containing ices 
and hydrocarbon species to form complex organic compounds
\citep[e.g][]{Foti:1984,Johnson:1984,Strazzulla:1991,McDonald:1996}. 

Laboratory VUV radiation experiments conducted on simple organic compounds in 
the presence of montmorillonite, kaolinite and volcanic ash 
yield 2-3 times more complex compounds than experiments 
without a mineral substrate \citep{Simakov:2005}.  
Thus, while most ion bombardment experiments are typically conducted without
the use of mineral substrates, they may limit carbonization processes by trapping and 
protecting newly formed compounds from photolysis 
\citep{Bonner:1985,Simakov:2005}.  
The increasing abundance of micron sized carbon particles with 
increasing irradiation time, as indicated by our modeling and SEM imaging, 
suggests these particles are being 
synthesized during the pulsed laser
irradiation of the Fe-rich sample and are not simply contamination products.  
Even if the micron sized carbon 
grains in our irradiated Fe-rich samples are inorganic, 
these carbon rich grains have not been observed in previous 
space weathering experiments. Thus, the formation of these
particles may be supported by the catalytic properties of cronstedtite
in addition to those of {\it n}pFe$^0$ particles \citep{Britt:2014}. 

Most organic irradiation experiments study the evolution of 
simple carbon species and/or organic-rich compounds that progress
towards more complex
organics, but as of yet, the production of organics from refractory materials 
has not been observed.   Although we cannot yet rule out the possibility of 
having synthesized some inorganic form of carbon, the following 
discussion examines the viability of organic synthesis in response to 
pulsed laser irradiation and its implications for the production of organics
on the surfaces of asteroids. 

Industrial Fischer-Tropsch (FT) type reactions commonly use iron oxides and 
metallic iron as catalysts for organic
synthesis (e.g. hydrocarbon fuels) \cite{Smit:2008,Britt:2014}.  In its simplest form, 
FT reactions combine CO and H$_2$ to form hydrocarbons and water, e.g. Eq. \ref{FTS}
\begin{equation} \label{FTS}
{\it n}CO + (2n+1)H_2 \rightarrow C_nH_{2n+2}+{\it n}H_2O
\end{equation}
The production of npFe (be it metallic or oxidized) from the
pulsed laser irradiation of our Fe-rich samples may provide a  
catalytic source for organic synthesis during our experiments.  
Additionally, studies on the thermodynamic properties of siderite show that for 
a range of pressures and temperatures, siderite decomposes to magnetite, 
graphite and CO$_2$ before melting \citep{Kang:2015}.  
Although our experiments are conducted at pressures much lower 
than the studies of \cite{Kang:2015},
we postulate that a similar type of decomposition produces feedstock carbon species 
during the irradiation of siderite.  Lastly, because our samples are not pressed into 
pellets (to better simulate asteroid-like regoliths), air trapped in our samples 
provide a source of N that would otherwise not be readily available in prior 
experiments. Thus, the presence of C, Fe-oxides (and/or Fe-metals), H$_2$O and N
provide the Fe-rich samples with an ample set of feedstock minerals and 
volatiles to potentially enable organic synthesis.  

Recent observations of ammoniated clays on Ceres \citep{de-Sanctis:2015} and 
ammonia contained within CCs \citep{Pizzarello:2012} suggest asteroids have 
notable reservoirs of N-bearing feedstock volatiles that are available for organic synthesis. 
Additional analyses are currently being conducted to 
better assess the nature of the micron sized carbon particles in our samples.  
These particles may represent the first step in a series of reactions that 
lead to complex, biologically significant 
organic compounds on the surfaces of asteroids \citep{Britt:2014}.  

\subsection{Comparison to space weathering experiments on meteorites}
The spectral bluing observed in this study appears to result from the production of 
large, micron sized organic-like particles, however, previous studies 
have found several other mechanisms that result in spectral bluing.  In particular, 
\cite{Moroz:1998} and \cite{Moroz:2004} find that ion bombardment of pure complex organics
results in spectral bluing due to brightening at UV wavelengths.  
\cite{Moroz:2004} conclude that the brightening in these experiments results form
carbonization, which increases the absorption coefficients and optical density of organics. 
Similar to the behavior of irradiated organics, 
both ion bombardment and pulsed-laser irradiation experiments on 
the meteorites Tagish Lake and Mighei 
produce flatter spectra (bluing) and brightening at UV wavelengths
(Table \ref{tab_pub_sw_results}).  Thus, the bluing of these meteorites is largely attributed
to the carbonization of organics \citep{Hiroi:2004,Vernazza:2013,Lantz:2015a}. 

\cite{Matsuoka:2015} also observes 
brightening and bluing of laser irradiated Murchison, but in addition to carbonization, 
they propose that the 10-1000nm Fe and S-rich vapor deposits also contribute 
to bluing.  In contrast, nanophase FeS particles are present in Itokawa samples \citep{Noguchi:2011}, 
and space weathering trends on this asteroid show spectral reddening\citep{Ishiguro:2007}. 
EDX analysis of our irradiated Fe-rich samples revealed 
small amounts of sulfur associated with the {\it n}pFe in the rims surrounding the 
carbon-rich melt grains.  The initial spectral reddening of the Fe-rich samples was only 
reproducible in our radiative transfer models through the use of {\it n}pFe$^0$ particles. 
Thus, just as \cite{Keller:2013} postulated, we conclude that
nanophase FeS particles likely have analogous optical 
effects as {\it n}pFe$^0$ particles.  
 
Although the {\it n}pFeS particles likely cause spectral reddening, 
another possible mechanism for spectral bluing may 
arise from larger Fe- and/or FeS-rich particles.  In addition to the 
$\mu$pC particles, $\sim$30-50 nm 
sized Fe and S-rich melt particles were observed in the irradiated Fe-rich samples. 
Although our radiative transfer models did not consider sulfur particles,  
the models that combine both small and large
Fe particles ({\it n}pFe$^0$ and $\mu$pFe$^0$) produce similar, 
albeit poorer fits to the final irradiated Fe-rich sample spectra 
as the best fit {\it n}pFe$^0$ and $\mu$pC models.  Thus, assuming 
that these larger particles behave similar to the modeled 
$\mu$pFe$^0$ particles, Fe and S-rich particles may also contribute 
to the bluing effect observed in our samples.  Nonetheless, our TEM data 
and radiative transfer models suggest that the $\mu$pC particles are 
an important and dominant irradiation product in this study. 

\subsection{Comparison to space weathering trends on volatile-rich asteroids}
The production of micron sized carbon grains has not yet been observed in other experiments, 
but presents a new mechanism for bluing that may have important implications regarding 
the spectral trends of asteroids.  
However, to date astronomical observations have predominantly shown that 
volatile-rich asteroids experience spectral reddening as a function of exposure age
\citep{Lazzarin:2006,Kaluna:2016,Fornasier:2016,Jaumann:2016}.  
High resolution images of Ceres obtained with the Dawn spacecraft \citep{Russell:2011} 
show a progression from young, blue impact craters to redder, 
older surface regions \citep{Jaumann:2016}.  
Ceres's mineralogy is dominated by Mg-rich serpentines and 
carbonates \citep{de-Sanctis:2015,Ammannito:2016}, which is
an indicator that the aqueous alteration on Ceres was progressive 
\citep{Howard:2009,Howard:2011,Rubin:2007a}. 
The spectral reddening observed in our Mg-rich lizardite samples 
are consistent with the space weathering trends observed on Ceres, 
and may indicate Mg-rich clays dominate the space weathering 
trends of this dwarf planet.  Similar to Ceres, the C-complex Themis and Beagle asteroid families 
show a progression towards spectrally redder surfaces with increasing 
exposure age \citep{Kaluna:2016,Fornasier:2016}, which may indicate that 
these asteroids are dominated by Mg-rich phyllosilicates. 
Although Dawn data shows evidence of extensive aqueous alteration, 
water ice and sublimation activity has also been detected on Ceres
\citep{Kuppers:2014,Combe:2016}.  Interestingly, the Themis and Beagle 
families are also associated with ice in the form of main-belt comets 
\cite[e.g.][]{Hsieh:2009,Novakovic:2012}.  This is further evidence of the 
similarities between Ceres and the Themis and Beagle asteroids, and 
indicates that prior to its disruption, the Themis/Beagle parent body may 
have had a similar composition and aqueous history as Ceres. 

While spectral reddening have been observed on Themis, Beagle
and Ceres, these trends may not be representative of those that occur on other volatile-rich 
asteroids.  \cite{Noguchi:2016} find that organics are lost with more pervasive aqueous alteration,
and thus bluing trends induced by carbonization may 
be associated with less altered C-complex asteroids. 
The Fe-rich sample used in this study is dominated by Fe-rich mineral cronstedtite, which is 
one of the dominant phyllosilicates formed during 
the early stages of aqueous alteration \citep{Howard:2009,Howard:2011,Rubin:2007a}.  
Thus, the spectral bluing trends 
of this sample may be representative of asteroids 
that have experienced small degrees of alteration.  
Additionally, \cite{Lantz:2015} found a broad correlation between the spectral evolution 
of meteorites and their initial albedos and possibly carbon content. 
While a much larger set of experiments are still needed to constrain the spectral trends 
of volatile-rich CCs, the data presented here along with the \cite{Lantz:2015} study 
appear to indicate that the compositional dependency of space weathering trends 
may be a useful tool to constrain the compositions and aqueous histories of volatile-rich asteroids.
An example of an asteroid family that may be inclined towards spectral bluing are the Veritas asteroids.
This family is noted for pervasive 0.7 $\mu$m features \citep{Kaluna:2015}, 
which are present in CM meteorites that are spectrally dominated by 
Fe-rich phyllosilicates \citep{Cloutis:2011}.  

\section{Conclusion}
We have shown that pulsed laser irradiation of the two sets of minerals formed during 
aqueously alteration result in notable but varied spectral variations.  
Irradiation of Mg-rich lizardite results in little slope and albedo variations, but significant 
absorption band variations in which hydroxyl features are less affected by laser irradiation 
than Fe features.  Irradiation of the Fe-rich sample (cronstedtite, pyrite and siderite) 
induces both spectral reddening and bluing, darkening, and absorption band suppression.  
SEM and TEM analyses of the Fe-rich assemblage revealed
micron sized organic-like (CNO-rich) particles that appear to have been produced during 
irradiation and may be a new mechanism for spectral bluing.  In addition to providing 
insight into the potential compositional dependency of space weathering 
trends on volatile-rich asteroids, these data also indicate that 
space weathering may play an important role in the synthesis of organics 
on asteroid surfaces.  

\acknowledgements
We would like to say mahalo nui to Karen Meech of the Institute for Astronomy, 
Jeff Taylor and Eric Hellebrand of HIGP, and Tina Carvalho of the PBRC 
for their contribution to and support of this work. 
Additionally, we would like to acknowledge Peter Eschbach at Oregon State University, 
and Karen Bustillo and Chengyu Song at the Molecular Foundry, Lawrence Berkeley Lab 
for their help with data collection.
This work was supported by the National Aeronautics
and Space Administration through the NASA Astrobiology Institute under
Cooperative Agreement No. NNA04CC08A issued through the Office of Space
Science, by NASA Grant No. NNX07A044G.  
Work was conducted at the Molecular Foundry (National Center of Electron Microscopy, 
Lawrence Berkeley National Lab) under Project 3688 (PI Ishii) and 
was supported by the Office of Science, Basic Energy Sciences, U.S. Department of 
Energy under Contract No. DE-AC02-05CH11231. 

\clearpage

\begin{deluxetable}{llllll}
\tabletypesize{\small}
\tablewidth{0pt}
\tablecaption{Previously Published Space Weathering Experiments on Volatile-rich CC Meteorites \label{tab_pub_sw_results}}
\tablehead{
\colhead{\bf Meteorite }&  
\colhead{\bf Petrologic} &
\colhead{\bf Slope}&
\colhead{\bf Albedo} &
\colhead{\bf Method} &
\colhead{\bf Reference} \\
\colhead{}&
\colhead{\bf Grade} &
\colhead{\bf Changes}	&
\colhead{\bf Changes$\dag$} &
\colhead{} &
\colhead{}  
}
\startdata
Murchison (CM) & 2.5-2.7$^{1,2}$ & Blueing & Increase ($<0.5\mu$m) & Laser & \cite{Matsuoka:2015} \\
Murchison  & 2.5-2.7 & NV & Decrease & He$^+$ Ion  & \cite{Lantz:2015} \\
Murchison  & 2.5-2.7 & NV & Increase ($<0.72\mu$m)  & Ar$^+$ Ion & \cite{Lantz:2015} \\
Murray (CM) & 2.4-2.6$^{1,2}$ & NQ & Decrease & H$^+$ Ion &  \cite{Hapke:1966} \\
Mighei (CM) & 2.3$^3$ & Reddening & NQ & Laser & \cite{Moroz:2004a}\\
Mighei  & 2.3 & Blueing & Increase &  He$^+$ Ion & \cite{Lantz:2015a}\\
Tagish Lake (C$^{\S}$) & 2.0$^{1,4}$  & Blueing & Increase & Laser & \cite{Hiroi:2004} \\
Tagish Lake  & 2.0 & Reddening & Increase & He$^+$ Ion & \cite{Vernazza:2013} \\
Tagish Lake  & 2.0 & Blueing & Increase ($<1\mu$m) & Ar$^+$ Ion & \cite{Vernazza:2013} \\
\enddata
\tablecomments{NV: No variations, NQ: Not quatified. 
Petrologic grades derived by: 
$^1$\cite{Rubin:2007}, 
$^2$\cite{Trigo:2006},
$^3$\cite{Rubin:2007a}, 
$^4$\cite{Gounelle:2001}.
$\dag$When provided, wavelength region where brightening occurs is given, 
$\S$A unique meteorite with affinities to CM and CI chondrites 
\citep{Zolensky:2002,Cloutis:2012c}. 
}
\end{deluxetable}
 
\begin{deluxetable}{llllllll} 
\tabletypesize{\small}
\tablecolumns{8}
\tablewidth{0pt}
\tablecaption{Microprobe Results \label{tab_microprobe}}
\tablehead{
\colhead{\bf Samples }&
\colhead{\bf SiO$_2$ }&  
\colhead{\bf FeO}&
\colhead{\bf MgO}&
\colhead{\bf Al$_2$O$_3$} &
\colhead{\bf SO$_3$}  &
\colhead{\bf CO$_2$}  &
\colhead{\bf H$_2$O$^{\ddag}$} 
}
\startdata
Olivine	& 41.13	& 9.51 	& 49.26 	&  -- 		&  -- 		& -- 		&   --         \\
Lizardite	& 39.34	& 1.68	& 39.71	& 0.22	& --   		& -- 		& 12.91  \\
\hline
{\bf Fe-rich assemblage} &&&&&&& \\
Cronstedtite	& 17.38	& 76.32$^{a}$  & -- 	& -- 	& 0.09 	& -- 	& 9.79 \\
Pyrite$^{b}$	& --		& 46.98		& --	& --	& 50.43	& --	& -- \\
Pyrite$^{bc}$	& 2.31	& 54.56		& --	& --	& 0.15	& --	& -- \\
Siderite	& --		& 60.68	& --		& -- 		& 0.03 	& 37.77 	& -- 
\enddata
\tablecomments{ Fractional abundances (wt.\%) determined from electron 
microprobe analyses of the minerals used in this work.  
$^{\ddag}$ H$_2$O abundance determined from stoichiometry. 
$^{a}$Values reflect Fe$_2$O$_3$ abundance. 
$^{b}$Values reflect elemental Fe and S abundances.
$^{c}$Oxidized pyrite phase. 
}
\end{deluxetable}

\begin{deluxetable}{llllllll}
\tabletypesize{\small}
\tablewidth{0pt}
\tablecaption{Average chemical compositions of Fe-rich assemblage melts  \label{tab_edx}}
\tablehead{
\colhead{\bf Sample}&  
\colhead{\bf Facility} &
\colhead{\bf Si}&
\colhead{\bf Fe} &
\colhead{\bf S} &
\colhead{\bf C} &
\colhead{\bf N} &
\colhead{\bf O}	
}
\startdata
FIB section 1		& OSU & 0.2$\pm$0.0 & 0.4$\pm$0.0 & {\it nd} & 66.1$\pm$2.1 & 17.0$\pm$0.7 &16.2$\pm$0.6   \\
FIB section 1	$^{\dag}$ & OSU & 10.0$\pm$2.5 & 55.9$\pm$11.4  & 8.2$\pm$4.9 & {\it nd} & {\it nd} &25.9$\pm$6.7 \\ 
FIB section 1		& NCEM & {\it nd} & {\it nd} & {\it nd} & 80.9$\pm$1.0 & 11.1$\pm$1.5 & 80.0$\pm$1.4 \\
FIB section 2		& NCEM 	&  {\it nd} &  {\it nd} & {\it nd} & 77.5$\pm$3.0 & 12.5$\pm$3.7 & 9.9 $\pm$4.2 \\
FIB section 2$^{\ddag}$	& NCEM & 5.3$\pm$3.8 & 45.5$\pm$7.8 & 4.3$\pm$6.7 & 15.2$\pm$2.0 & 4.4$\pm$3.1 & 25.3$\pm$2.3 \\
\enddata
\tablecomments{ Abundances are expressed in wt.\%. $^{\dag}$Average of two adjacent Fe melt regions $^{\ddag}$Composition of rim surrounding carbon rich grain. 
}
\end{deluxetable}

\begin{center}
\begin{deluxetable}{cccc}
\tabletypesize{\small}
\tablewidth{0pt}
\tablecaption{Best fit radiative transfer model
abundances  \label{tab_model_abundances}}
\tablehead{
\colhead{\bf Irradiation}&  
\colhead{\bf Nanophase Iron} &
\colhead{\bf Microphase Carbon} &
\colhead{\bf $\chi^2_{red}$}	 
\\
\colhead{\bf Time}&
\colhead{\bf Abundance (wt\%)}&
\colhead{\bf Abundance (wt\%)}&
\colhead{\bf }
}
\startdata
{\bf Olivine$^a$} && &\\
\hline
2.5 min &  0.003 & -- &1.004  \\
5.0 min &  0.008 & --&1.070  \\
10 min &  0.018 & -- & 4.101 \\
15 min &  0.028 & -- &1.961 \\
20 min &  0.043 & -- &4.357 \\
30 min &  0.064 &  -- &1.594 \\
40 min &  0.090 & -- &12.495  \\
\hline
{\bf Lizardite$^b$} &&& \\
\hline
2.5 min &  0.000 & -- &0.357  \\
5.0 min &  0.001 & -- &0.423 \\
10 min &  0.002 &  -- &0.938   \\
15 min &  0.002 &--  &2.239  \\
20 min &  0.004 & -- & 1.297   \\
30 min &  0.006 &-- & 2.629   \\
40 min &  0.003 &-- & 7.543  \\
\hline
{\bf Fe-rich assemblage$^c$} && &\\
\hline
2.5 min & 0.078 & 0.028 & 1.646 \\ 
5.0 min & 0.162 & 0.053 & 1.646  \\
10 min & 0.223 & 0.103 & 1.646 \\
15 min & 0.190 & 0.122 & 1.646 \\
20 min & 0.155 & 0.189 & 3.241 \\
30 min & 0.175 & 0.312 & 3.241 \\
40 min & 0.000 & 0.417 & 3.241 \\
\enddata
\tablecomments{ Each of the models were produced using mean grain diameters of 37.5$\mu$m. 
$^a$Olivine models were produced using a grain density ({\it $\rho$}) of 3.32 g/cm$^3$ and an index of refraction ({\it n}) of 1.64. $^b$Lizardite models were produced using {\it $\rho$} = 2.55 g/cm$^3$ and {\it n} = 1.60. $^c$Lizardite models were produced using {\it $\rho$} = 3.34 g/cm$^3$ and {\it n} = 1.72.}
\end{deluxetable}
\end{center}

\clearpage

\clearpage

\clearpage

\bibliographystyle{model5-names}

\bibliography{SW_AqAltMinerals.bib}

\end{document}